\newcommand{\poly}{\tilde{T}(z)}
\newcommand{\tr}{{\rm tr}}
\newcommand{\diag}{{\rm diag}}

\newcommand{\up}{u'} 
\newcommand{\hA}{\tilde{A}} 
\newcommand{\hB}{\tilde{B}}
\newcommand{\hM}{\tilde{M}}
\newcommand{\hN}{\tilde{N}}
\newcommand{\hP}{\tilde{P}}
\newcommand{\hR}{\tilde{R}}
\newcommand{\hS}{\tilde{S}}
\newcommand{\hT}{\tilde{T}}

\newcommand{\hW}{\tilde{W}}
\newcommand{\ha}{\tilde{a}}
\newcommand{\hb}{\tilde{b}}
\newcommand{\he}{\tilde{e}}
\newcommand{\hh}{\tilde{h}}
\newcommand{\hi}{\tilde{\imath}}
\newcommand{\hj}{\tilde{\jmath}}

\newcommand{\hPhi}{\tilde{\Phi}}
\newcommand{\hPsi}{\tilde{\Psi}}
\newcommand{\hla}{\tilde{\lambda}}
\newcommand{\hrho}{\tilde{\rho}}
\newcommand{\hphi}{\tilde{\phi}}

\newcommand{\hLap}{\tilde{\Lambda}'}

\newcommand{\home}{\tilde{\omega}}
\newcommand{\tx}{\tilde{x}}
\newcommand{\tX}{\tilde{X}}
\newcommand{\tY}{\tilde{Y}}

\newcommand{\bC}{\bar{C}}

\newcounter{abc}

\def\pre{ \cF(a) }
\def\pree{ \cF(a, \ha) }
\def\mS{ \bar{S} }
\def\laSW{ \lambda_{SW} }
\def\gs{ g_{s} }
\def\fre{ F_{\rm s} }
\def\fren{ F^{(n)}_{\rm s} }
\def\frethree{ F^{(3)}_{\rm s} }
\def\free{ \fre(S, \mS) }
\def\freen{ \fren(S, \mS) }

\def\freethree{ \frethree(S, \mS) }
\def\frpp{ F_{\rm rp}  }
\def\frp{\frpp (S) }
\def\frpn{ F^{(n)}_{\rm rp} (S) }
\def\frptwo{ F^{(2)}_{\rm rp}(S) }
\def\Weff{ W_{\rm eff}}
\def\eij{  e_{ij} }
\def\eik{  e_{ik} }

\def\eil{  e_{i\ell} }
\def\ekl{  e_{k\ell} }
\def\heij{ \he_{ij} }

\def\hekl{ \he_{k\ell} }
\def\gij{  g_{ij} }
\def\gik{  g_{ik} }
\def\gjk{  g_{jk} }

\def\gil{  g_{i\ell} }
\def\gkl{  g_{k\ell} }
\def\glk{  g_{\ell k} }
\def\hij{  h_{ij} }
\def\hhij{ \hh_{ij} }
\def\hik{  h_{ik} }

\def\hjk{  h_{jk} }
\def\hji{  h_{ji} }
\def\hil{  h_{i\ell} }

\def\hkl{  h_{k\ell} }
\def\hlk{  h_{\ell k} }
\def\aij{  a_{ij} }
\def\aik{  a_{ik} }
\def\ajk{  a_{jk} }

\def\ail{  a_{i\ell} }
\def\Gi{ G_i}
\def\Gj{ G_j}
\def\Gk{ G_k}
\def\Gl{ G_\ell}
\def\Ti{ T_i}
\def\Tj{ T_j}
\def\Tk{ T_k}
\def\Tl{ T_\ell}
\def\hTi{ \hT_i}
\def\hTj{ \hT_j}
\def\hTk{ \hT_k}
\def\hTl{ \hT_\ell}

\def\sumN{ \sum_{i=1}^{N} }
\def\sumjN{ \sum_{j=1}^{N} }
\def\prodN{ \prod_{i=1}^{N} }
\def\sumk{ \sum_{k\neq i} }
\def\suml{ \sum_{\ell \neq i,k} }

\def\vev#1{ \langle {#1} \rangle }

\def\bvev#1{ \bigg\langle {#1} \bigg\rangle }

\def\tadpole{ \vev{\tr\, \Psi_{ii} }_{S^2} }
\def\htadpole{ \vev{\tr\, \hPsi_{ii} }_{S^2} }
\def\rptadpole{ \vev{\tr\, \Psi_{ii} }_{\sR \sPP^2}   }

\def\sphere{ \bigg|_{\rm sphere} }
\def\rp{ \bigg|_{\sR \sPP^2}   }
\def\vevS{ \bigg|_{\vev{S}}  }
\def\vevShS{ \bigg|_{\vev{S},\vev{\hS}} }

\newcommand{\be}{\begin{equation}}
\newcommand{\ee}{\end{equation}}
\newcommand{\bea}{\begin{eqnarray}}
\newcommand{\eea}{\end{eqnarray}}
\newcommand{\ba}{\begin{array}}
\newcommand{\ea}{\end{array}}

\newcommand{\al}{\alpha}

\newcommand{\de}{\delta}

\newcommand{\si}{\sigma}
\newcommand{\om}{\omega}

\newcommand{\lam}{\lambda}
\newcommand{\la}{\lambda}
\newcommand{\La}{\Lambda}
\newcommand{\Lap}{\La'}
\newcommand{\Ups}{\Upsilon}

\newcommand{\half}{{\textstyle {1\over 2}}}
\newcommand{\third}{ {\textstyle{1\over 3}}}
\newcommand{\fourth}{ {\textstyle{1\over 4}}}

\newcommand{\x}{z}
\newcommand{\Qx}{P(-\x)}
\newcommand{\LU}{\Lambda^N}
\newcommand{\LUp}{\Lambda'^N}
\newcommand{\LA}{\Lambda^{N+2}}
\newcommand{\LAp}{\Lambda'^{N+2}}
\newcommand{\LS}{\Lambda^{N-2}}
\newcommand{\LSp}{\Lambda'^{N-2}}
\newcommand{\LX}{\Lambda^{N-2\beta}}
\newcommand{\LXp}{\Lambda'^{N-2\beta}}

\newcommand{\PP}{\mathrm{I}\kern -2.5pt \mathrm{P}}
\newcommand{\R}{\mathrm{I}\kern -2.5pt \mathrm{R}}
\newcommand{\sPP}{\mathrm{I}\kern -1.6pt \mathrm{P}}
\newcommand{\sR}{\mathrm{I}\kern -1.6pt \mathrm{R}}
\newcommand{\Z}{\mathsf{Z}\kern -5pt \mathsf{Z}}
\newcommand{\C}{\mathsf{I}\kern -5pt \mathrm{C}}

\newcommand{\e}{{\rm e}}
\newcommand{\D}{{\rm d}}
\newcommand{\pa}{\partial}
\newcommand{\rar}{\rightarrow}
\newcommand{\non}{\nonumber}

\newcommand{\cN}{\mathcal{N}}
\newcommand{\cW}{\mathcal{W}}
\newcommand{\cF}{\mathcal{F}}
\newcommand{\cO}{\mathcal{O}}

\newcommand{\ts}{\textstyle}

\newcommand{\1}{1\kern -3pt \mathrm{l}}

\newcommand{\U}{\mathrm{U}}
\newcommand{\UxU}{\U(N){\times}\U(N)}
\newcommand{\SO}{\mathrm{SO}}
\newcommand{\Sp}{\mathrm{Sp}}

\newcommand{\so}{\mathrm{so}}
\newcommand{\spl}{\mathrm{sp}}

\documentclass[12pt]{article}

\usepackage{graphicx}

\def\theequation{\thesection.\arabic{equation}}

\newdimen\tableauside\tableauside=1.0ex
\newdimen\tableaurule\tableaurule=0.4pt
\newdimen\tableaustep
\def\phantomhrule#1{\hbox{\vbox to0pt{\hrule height\tableaurule
width#1\vss}}}
\def\phantomvrule#1{\vbox{\hbox to0pt{\vrule width\tableaurule
height#1\hss}}}
\def\sqr{\vbox{%
  \phantomhrule\tableaustep

\hbox{\phantomvrule\tableaustep\kern\tableaustep\phantomvrule\tableaustep}%
  \hbox{\vbox{\phantomhrule\tableauside}\kern-\tableaurule}}}
\def\squares#1{\hbox{\count0=#1\noindent\loop\sqr
  \advance\count0 by-1 \ifnum\count0>0\repeat}}
\def\tableau#1{\vcenter{\offinterlineskip
  \tableaustep=\tableauside\advance\tableaustep by-\tableaurule
  \kern\normallineskip\hbox
    {\kern\normallineskip\vbox
      {\gettableau#1 0 }%
     \kern\normallineskip\kern\tableaurule}%
  \kern\normallineskip\kern\tableaurule}}
\def\gettableau#1 {\ifnum#1=0\let\next=\null\else
  \squares{#1}\let\next=\gettableau\fi\next}

\newcommand{\Yfund}{\tableau{1}}
\newcommand{\Ysymm}{\tableau{2}}
\newcommand{\Yasymm}{\tableau{1 1}}

\begin{document}

\begin{flushright} 
{\tt hep-th/0305263}\\ 
BRX-TH-519\\ 
BOW-PH-129 \\
\end{flushright}
\vspace{1mm} 
\begin{center}
{\bf\Large\sf Matrix-model description of {\large $\cN=2$ } gauge theories
 \\ with non-hyperelliptic Seiberg-Witten curves } 

\vskip 5mm

Stephen G. Naculich\footnote{Research
supported in part by the NSF under grant PHY-0140281.}$^{,a}$,
Howard J. Schnitzer\footnote{Research
supported in part by the DOE under grant DE--FG02--92ER40706.}$^{,b}$,
and Niclas Wyllard\footnote{Research 
supported by the DOE under grant DE--FG02--92ER40706.\\
{\tt \phantom{aaa} naculich@bowdoin.edu;
schnitzer,wyllard@brandeis.edu}\\}$^{,b}$

\end{center}

\vskip 1mm

\begin{center}
$^{a}${\em Department of Physics\\
Bowdoin College, Brunswick, ME 04011}

\vspace{.2in}

$^{b}${\em Martin Fisher School of Physics\\
Brandeis University, Waltham, MA 02454}

\end{center}

\vskip 2mm

\begin{abstract} 

Using matrix-model methods we study three different 
$\cN=2$ models: $\U(N){\times}\U(N)$ 
with matter in the bifundamental representation,
$\U(N)$ with matter in the symmetric representation, and 
$\U(N)$ with matter in the antisymmetric representation. 
We find that the (singular) cubic Seiberg-Witten curves 
(and associated Seiberg-Witten differentials) implied by the matrix models, 
although of a different form 
from the ones previously proposed using M-theory, can be transformed 
into the latter and are thus physically equivalent. 
We also calculate the one-instanton corrections to 
the gauge-coupling matrix using the perturbative
expansion of the matrix model. For the $\U(N)$ theories 
with symmetric or antisymmetric matter
we use the modified matrix-model 
prescription for the 
gauge-coupling matrix discussed in ref.~\cite{Naculich:2003}.
Moreover, in the matrix model for the $\U(N)$ theory with antisymmetric matter,
one is required to expand around a different vacuum 
than one would naively have anticipated.  
With these modifications of the matrix-model prescription,
the results of this paper are in complete agreement
with those of Seiberg-Witten theory obtained using M-theory methods.

\end{abstract}

\setcounter{equation}{0}
\section{Introduction}

The study of $\cN=2$ gauge theories using the matrix-model 
techniques of Dijkgraaf and Vafa was initiated 
in~\cite{Dijkgraaf:2002b,Dijkgraaf:2002d} (see \cite{Cachazo:2002a} 
for earlier work) and was further 
developed and studied 
in~\cite{Naculich:2002a,Klemm:2002,Naculich:2002b,Matone:2002,
Ahn:2003a,Abbaspur:2003} (see also \cite{Boels:2003}). 
In \cite{Naculich:2002b} we showed that it is possible to 
derive all the building blocks of the Seiberg-Witten (SW)
solution \cite{Seiberg:1994a} 
(i.e.~the curve and a preferred meromorphic differential) 
purely within the matrix-model context. 
Since the curve is obtained from the 
large-$M$ solution of the matrix model, 
one can obtain the SW curve in this manner 
only when an explicit large-$M$ solution is available.
However, as was stressed in \cite{Dijkgraaf:2002d,Naculich:2002a},
even when the large-$M$ solution is not available, 
one can still resort to perturbation theory to derive the 
prepotential order-by-order without knowledge of 
the curve or differential. 
A necessary ingredient for this procedure
is the knowledge of the quantum order parameters $a_i$ 
(periods of the SW curve); in~\cite{Naculich:2002a}  
we proposed a perturbative definition of the periods.

In this paper, we extend the matrix-model program to   
 three $\cN=2$ gauge theories 
whose SW curves are non-hyperelliptic:
$\U(N){\times}\U(N)$ with a bifundamental hypermultiplet,
$\U(N)$ with a symmetric hypermultiplet,
and $\U(N)$ with an antisymmetric hypermultiplet. In sec.~\ref{sMtheory} 
we review some previous results for these theories 
and reformulate them in a way that will facilitate comparison with our 
later discussion.

The equivalence of the $\cN=1$ versions of the gauge theories 
considered in this paper and the 
corresponding matrix models was established, following the approach 
in \cite{Cachazo:2002b}, in refs.~\cite{Naculich:2003,Casero:2003} 
(related earlier and subsequent developments can be found 
in~\cite{Dijkgraaf:2002e,Seiberg:2002,
Kraus:2003a,Alday:2003,Kraus:2003b,Aganagic:2003}).
In principle these general results also show the validity of 
the matrix-model approach in the $\cN=2$ limit.
However, to obtain precise information about the $\cN=2$ gauge theories 
requires substantial additional work. Furthermore, there are several 
subtle issues, particularly for the 
theory with antisymmetric matter, that need to be addressed in order 
to recover the known results from the matrix models.

In sec.~\ref{sPert} we compute the gauge-coupling matrices $\tau_{ij}$
using the perturbative expansion of the matrix model.
For the $\U(N){\times}\U(N)$ model,  the gauge-coupling 
matrices are given by the second derivative of 
the free energy of the matrix 
model \cite{Dijkgraaf:2002c,Dijkgraaf:2002e,Cachazo:2002b}.
However, for the $\U(N)$ theories with symmetric or antisymmetric 
hypermultiplets, 
certain crucial signs must be included 
among the terms of the second derivative of the free energy 
to obtain $\tau_{ij}$.
The rationale for this was described in ref.~\cite{Naculich:2003},
and is implemented in sec.~\ref{sSAperttau}.

Furthermore, for the $\U(N)$ theory with antisymmetric matter,
we will show in sec.~\ref{sPertSA} that,
in order to obtain a gauge-coupling matrix 
and $\cN=2$ prepotential
that agree with those computed using SW theory,
one must perturbatively expand the matrix model
around a vacuum different from the one that
would have been naively anticipated.
In subsequent sections, we will see that this
choice of vacuum is similarly required to 
reproduce the known SW curve and differential.
The underlying reason for this choice of vacuum
remains somewhat obscure to us.

The matrix models associated with the gauge theories
we consider can be solved in the large-$M$ limit, 
giving rise to cubic algebraic 
curves~\cite{Dijkgraaf:2002b,Hofman:2002,
Lazaroiu:2003,Klemm:2003,Naculich:2003}
\be
u^3 - r(z) \, u - s(z) = 0 \,.
\ee
The functions $r(z)$ and $s(z)$ 
are determined by the matrix-model potential, 
up to two arbitrary polynomials $r_1(z)$ and $t_1(z)$
(which can be expressed in terms of the eigenvalues 
of the adjoint field(s) 
of the matrix model \cite{Lazaroiu:2003,Klemm:2003,Naculich:2003}).
To fix the forms of $r_1(z)$ and $t_1(z)$,
one must impose an additional criterion,
namely, the extremization of the effective superpotential
of the associated gauge theory.

In sec.~\ref{sMMcurve},
we use matrix-model perturbation theory
to provide a simple and efficient method
for determining the polynomials $r_1(z)$ and $t_1(z)$,
and thus the cubic curve,
in an expansion in the quantum scale $\La$.
We show that the cubic curves obtained 
for each of the theories considered
can be transformed into the SW curves of those theories 
obtained using M-theory (and also from geometric engineering 
for the $\U(N){\times}\U(N)$ theory).

Also in sec.~\ref{sMMcurve},
we use extremization of $\Weff$ together with the
saddle-point solution to derive a condition which 
implies, via Abel's theorem, the existence of a 
certain function on the matrix-model curve.
We then show that such a function exists on the known
(exact) M-theory curves.
Assuming uniqueness,
this demonstrates that extremization of $\Weff$ 
leads to a matrix-model curve that agrees
exactly with the M-theory curve.

The Seiberg-Witten differentials for the gauge theories 
studied in this paper can also be obtained within the 
matrix-model framework.   
In sec.~\ref{sSWdiff},
we compute these using matrix-model perturbation theory,
obtaining agreement with the SW differentials known from
M-theory.

Appendix A contain a derivation of the SW curve and differential 
for the $\cN=2$ $\U(N)$ theory with fundamental hypermultiplets
using methods developed in sec.~\ref{sMMcurve} and \ref{sSWdiff}
of this paper.
Appendix B contains some technical details of 
the calculations of section \ref{sPert}.

\setcounter{equation}{0}
\section{Cubic Seiberg-Witten curves from M-theory}
\label{sMtheory}

The Seiberg-Witten curves and differentials for the
$\cN=2$ gauge theories considered in this paper 
were previously obtained using M-theory methods, 
following the approach of ref.~\cite{Witten:1997}. 
(The $\UxU$ curve can also 
be obtained using geometric engineering~\cite{Katz:1997}.) 

The SW curves for the theories: \\[4pt]
\indent (a) \quad $\cN=2$ $\UxU$ with an $\cN=2$ 
bifundamental hypermultiplet \cite{Witten:1997,Erlich:1998}, \\[4pt]
\indent (b) \quad $\cN=2$ $\U(N)$ model with one 
symmetric ($\Ysymm$) hypermultiplet \cite{Landsteiner:1998a}, \\[4pt]
\indent (c) \quad $\cN=2$ $\U(N)$ model with one 
antisymmetric ($\Yasymm$) hypermultiplet \cite{Landsteiner:1998a}, \\[4pt]
obtained from M-theory considerations are given by
\renewcommand{\theequation}{\arabic{section}.\arabic{equation}\alph{abc}}
\setcounter{abc}{1}
\bea \label{U2Mcurve}
&& y^3 + P(\x)\, y^2 + \Lap^N \, \hP(\x) \,y 
+ (\Lap\,^2 \hLap)^N = 0  \,,
\\[8pt]\addtocounter{equation}{-1}\addtocounter{abc}{1}  \label{USMcurve}
&& y^3 + P(\x)\, y^2 + \Lap^{N-2} \x^2 \Qx \, y 
+ \Lap\,^{3N-6} \x^6 = 0\,, 
\\[3pt] \addtocounter{equation}{-1}\addtocounter{abc}{1} \label{UAMcurve}
&& y^3 + 
\bigg[P(\x) + {3\Lap^{N+2} \over \x^2} \bigg] y^2 
+ {\Lap^{N+2}\over \x^2}
 \bigg[\Qx + {3 \Lap^{N+2} \over \x^2} \bigg] y +{ \Lap\,^{3N+6}\over \x^6}
 =0 \,,
\eea
\renewcommand{\theequation}{\arabic{section}.\arabic{equation}}where
$P(\x) = \prod_{i=1}^N (\x-e'_i)$ and $\hP(\x) = \prod_{i=1}^N (\x-\he'_i)$.
The Seiberg-Witten differential for each of the theories 
above, obtained from the M-theory setup, is given 
by~\cite{Witten:1997,Fayyazuddin:1997,Landsteiner:1998a}
\be
\label{laSWm}
\laSW = \x \, \frac{\D y }{y} \,.
\ee

The map $y \rar {(\Lap \hLap)^N /y}$ 
in the curve (\ref{U2Mcurve}) 
corresponds to exchanging the two factors of the gauge group, 
i.e.~it can be undone by interchanging $e'_i \leftrightarrow \he'_i$, 
$\hLap \leftrightarrow \Lap$, and thus leads 
to a physically equivalent curve.
The two curves (\ref{USMcurve}), (\ref{UAMcurve}) are 
invariant under the involutions
$\x \rar -\x$, $y \rar {\Lap\,^{2N-4}\x^4/y}$, and 
$\x \rar -\x$, $y \rar {\Lap\,^{2N+4}/\x^4 y}$, respectively. 
The actual SW curves for these theories
are the quotients of the curves 
(\ref{USMcurve}), (\ref{UAMcurve}) 
by the involution.
This reflects the presence of the orientifold plane in the
type IIA brane configurations that lift to the M-theory configurations 
leading to these curves. 

Using (\ref{U2Mcurve})-(\ref{UAMcurve}) and (\ref{laSWm}), 
the leading term in the instanton expansion of the prepotential 
for each of the theories above was 
derived in refs.~\cite{Naculich:1998a,Ennes:1998b,Ennes:1998c}. 
Recently a more efficient 
method has been developed \cite{D'Hoker:2002,Gomez-Reino:2003}  
based on earlier work \cite{Chan:1999}. 
(See also~\cite{Nekrasov:2002} for another approach.)
In sec.~\ref{sPert}, we will reproduce these results from a 
perturbative matrix-model calculation.

For later comparison with matrix model results (sec.~\ref{sMMcurve}), 
we need to transform the above curves into another form,
which is invariant under the maps discussed above.
To do this we define\footnote{Henceforth, for simplicity, we will 
set $\hLap= \Lap$ in the $\UxU$ model.}
\bea
\label{defu}
\up &=& - y -2 \LXp \x^{2 \beta} - {\Lap\,^{2N-4\beta} \x^{4 \beta} \over y} \,,\\
\label{defw}
w &=& \half  \x^{-\beta} 
\left[y - {\Lap\,^{2N-4\beta} \x^{4 \beta} \over y} \right],
\eea
where $\beta=0$, $1$, $-1$ for curves 
(\ref{U2Mcurve}), (\ref{USMcurve}), and (\ref{UAMcurve}) respectively;
note that $\up$ is invariant under the maps discussed above.
The variables $\up$, $w$ are related via
\setcounter{abc}{1}
\renewcommand{\theequation}{\arabic{section}.\arabic{equation}\alph{abc}}
\bea \label{U2w2}
4 w^2 &=& \up^2  + 4 \LUp \up \,,
\\ \addtocounter{equation}{-1}\addtocounter{abc}{1} \label{USw2}
4 \x^2 w^2 &=& \up^2 + 4 \LSp \x^2 \up \,,
\\ \addtocounter{equation}{-1}\addtocounter{abc}{1} \label{UAw2}
4 w^2 &=& \x^2 \up^2 +  4 \LAp \up \,.
\eea
\renewcommand{\theequation}{\arabic{section}.\arabic{equation}}
Using eq.~(\ref{defu}) and eqs.~(\ref{U2Mcurve})-(\ref{UAMcurve}), 
one may show that
\setcounter{abc}{1}
\renewcommand{\theequation}{\arabic{section}.\arabic{equation}\alph{abc}}
\bea  \label{U2y}
\qquad y &=& - \LUp \left( \up - \hP(\x) + 3 \LUp \over  \up - P(\x) + 3 \LUp 
\right),
\\ \addtocounter{equation}{-1}\addtocounter{abc}{1} \label{USy}
\qquad y 
&=& -\LSp \x^2 \left( \up - \Qx + 3 \LSp \x^2 \over \up - P(\x) 
+ 3 \LSp \x^2 \right),
\\ \addtocounter{equation}{-1}\addtocounter{abc}{1} \label{UAy}
\qquad y &=& - {\LAp \over \x^2} \left( \up - \Qx\over \up - P(\x) \right).
\eea
\renewcommand{\theequation}{\arabic{section}.\arabic{equation}}Next, 
use the definition of $\up$ to write 
$w = \x^{-\beta} y + \half \x^{-\beta} \up + \LXp \x^{\beta}$.
Substitute eqs.~(\ref{U2y})-(\ref{UAy})  into this equation,
square it, and use eqs.~(\ref{U2w2})-(\ref{UAw2}) to find\footnote{The form 
of the curve for $\U(N)$ with one antisymmetric 
hypermultiplet (\ref{UAucurve}) 
was first obtained in sec.~7 of ref.~\cite{Landsteiner:1998a}, where the
connection to the Atiyah-Hitchin space (\ref{UAw2}) was also discussed.}
\setcounter{abc}{1}
\renewcommand{\theequation}{\arabic{section}.\arabic{equation}\alph{abc}}
\bea \label{U2ucurve} 
&& \!\!\!\!\! 
(\up - P(\x) + 3 \LUp)\,\up\,(\up - \hP(\x) + 3 \LUp)  
= \LUp (P(\x)-\hP(\x))^2 \,,
\\[5pt] \addtocounter{equation}{-1}\addtocounter{abc}{1} \label{USucurve}
&& \!\!\!\!\!
(\up - P(\x) + 3 \LSp \x^2)\,\up\,(\up - \Qx + 3 \LSp \x^2) 
 =  \LSp \x^2 (P(\x)-\Qx)^2 \,,
\\  \addtocounter{equation}{-1}\addtocounter{abc}{1} \label{UAucurve}
&& \!\!\!\!\! 
(\up - P(\x))\,\up\,(\up - \Qx)  =  {\LAp\over \x^2} (P(\x)-\Qx)^2 \,.
\eea
\renewcommand{\theequation}{\arabic{section}.\arabic{equation}}In
sec.~\ref{sMMcurve}, we will compare the matrix-model curves to
the SW curves written in this form. 

On the last two curves,
the involution acts as $\x \to -\x$, with $\up$ invariant.
The invariance of the curves under the involution
means that the equations can be written in terms
of $\up$ and $\x^2$.
The actual SW curve (quotient by the involution)
is thus a cover of the $\x^2$ plane. 
The first curve (\ref{U2ucurve}) 
is invariant under the interchange of the two gauge groups. 

Were we to reverse the transformation, 
described in the last paragraph, from the curves 
(\ref{U2Mcurve})-(\ref{UAMcurve}) 
to the ones in (\ref{U2ucurve})-(\ref{UAucurve}),
we would obtain two solutions, 
due to the fact that we squared both sides of an equation 
in one of the steps above. 
However, these two solutions are related by the involution
in cases (b) and (c).
In case (a), the two solutions are related by 
$e'_i \leftrightarrow \he'_i$, i.e. $P \leftrightarrow \hP$ and 
so correspond to exchanging the two $\U(N)$ factors. 
Hence in all cases the two solutions are physically equivalent. 

When $\Lap \to 0$, 
the curves (\ref{U2ucurve})-(\ref{UAucurve}) are singular 
at the roots of $P(z)$, $\hP(z)$ (or $P(-z)$), 
and $P(z)-\hP(z)$ (or $P(z)-P(-z)$).
(The discriminant has double zeros at those points.)
When $\Lap \neq 0$, the surface is deformed 
such that the first two sets of singular points 
open up into branch cuts, 
but the singularities at the points $z$ where
$P(z)= \hP(z)$ (or $P(z)=P(-z)$) 
remain\footnote{
This fact is important to get the genus counting to 
work, cf.~sec.~\ref{sMMcurve}. 
}.
One (important) exception occurs for the theory with antisymmetric matter, 
where the singularity at $z=0$ also opens 
up into a branch cut in the curve (\ref{UAucurve}).
For the theory with symmetric matter, $z=0$ remains a singular point 
in the curve (\ref{USucurve}).

\setcounter{equation}{0}
\section{Perturbative approach to the matrix model}
\label{sPert}

As we discussed in \cite{Naculich:2002a} 
(see also~\cite{Dijkgraaf:2002d}),
the $\cN=2$ gauge theory prepotential (in an instanton expansion)
may be determined using only matrix-model perturbation theory. 
In this approach, 
one adds to the $\cN=2$ superpotential an additional piece
which freezes the moduli to a generic, but fixed, 
point on the Coulomb branch of the $\cN=2$ theory,
and breaks the $\cN=2$ supersymmetry down to $\cN=1$.
After the relevant quantities are computed,
the extra piece is removed, restoring $\cN=2$ supersymmetry.  

This approach was first explored in \cite{Dijkgraaf:2002d} 
where the gauge-coupling matrix $\tau_{ij}$ was determined for $\U(2)$, 
and  was extended to $\U(N)$ in \cite{Naculich:2002a} where, in 
particular, a proposal for how to determine the relation between 
the quantum order parameters $a_i$ and their classical counterparts $e_i$ 
entirely within the context of matrix-model perturbation theory was put 
forward. 
Using this proposal the prepotential $\cF(a)$ was calculated 
to one-instanton level 
and was shown to agree with the well-known result.
In \cite{Naculich:2002b} the calculation was extended to include matter 
in the fundamental representation, 
and in~\cite{Ahn:2003a,Abbaspur:2003} to $\SO/\Sp$ gauge groups.

In this section, we extend this perturbative matrix-model method 
to new cases by calculating the one-instanton contribution
to the $\cN=2$ prepotential in the 
$\U(N){\times}\U(N)$ gauge theory with a bifundamental hypermultiplet,
and the $\U(N)$ gauge theory with one 
symmetric ($\Ysymm$) or antisymmetric ($\Yasymm$) hypermultiplet.
Besides the additional complication of dealing with two-index matter,
and the inclusion of diagrams with the topology 
of $\R\PP^2$ (for $\Ysymm$ and $\Yasymm$),
there is one 
significant modification of the procedure 
developed in refs.~\cite{Naculich:2002a, Naculich:2002b}: 
namely, for the models with symmetric or antisymmetric matter,
$\tau_{ij}$ is no longer given simply by the 
second derivative of the free energy \cite{Naculich:2003}
(see sec.~\ref{sSAperttau}). 
Moreover, in the matrix model for the $\U(N)$ theory with antisymmetric matter,
one is required to expand around a different vacuum 
than one would naively have anticipated (this may be related 
to the results in \cite{Oda:1998}).

Previously the one-instanton prepotential for these models
has been obtained using M-theory 
methods \cite{Naculich:1998a, 
Ennes:1998b, Ennes:1998c,D'Hoker:2002,Gomez-Reino:2003}
(see also \cite{Nekrasov:2002} for another approach).

\subsection{ $\U(N){\times}\U(N)$ with a bifundamental hypermultiplet} 
\label{sPertUU}

Consider the $\cN=1$ $\UxU$ 
supersymmetric gauge theory with the following matter content: 
two chiral superfields 
$\phi_i{}^j$, $\hphi_{\hi}{}^{\hj}$ transforming 
in
the adjoint representation of each of the two factors of the gauge group, 
one chiral superfield $b_i{}^{\hj}$ 
transforming in the bifundamental representation 
$(\Yfund,\overline{\Yfund})$, and 
one chiral superfield $\hb_{\hi}{}^j$ transforming in the 
bifundamental representation $(\overline{\Yfund},\Yfund)$.
The superpotential of this gauge theory is taken to be of the 
form\footnote{A mass term for the bifundamental fields, $m\, \tr(\hb \,b)$, 
can be introduced by 
shifting $\phi \rar  \phi - m/2$ ($\hphi \rar \hphi+m/2$) and 
$e_i \rar e_i - m/2$ ($\he_i\rar \he_i+m/2$).}
\be \label{U2suppot}
\cW(\phi,\hphi,b,\hb) = \tr \,[ W(\phi) - \hW(\hphi) - \hb \, \phi \,b 
+ b\, \hphi \,\hb ] \,,
\ee
where $W(\phi)$ and $\hW(\hphi)$ are polynomials such that 
\be\label{Wdef}
W'(z) = \al \prod_{j=1}^N (z-e_j), \qquad
\hW'(z) = \al \prod_{j=1}^N (z - \he_j)\,. 
\ee
The superpotential (\ref{U2suppot})
can be viewed as a deformation of an $\cN=2$ theory,
which is recovered when $\alpha \to 0$. At the end of our calculation 
we will take this limit thereby obtaining results 
valid in the $\cN=2$ theory.

The $\U(M){\times}\U(\hM)$ matrix model
associated with this gauge theory \cite{Dijkgraaf:2002b,
Hofman:2002, Lazaroiu:2003,Naculich:2003}
has partition function\footnote{We 
use capital letters to denote matrix model quantities.}
\be 
\label{ZUU}
Z = {1 \over {\rm vol~} G}
 \int \D \Phi \, \D \hPhi \, \D B \, \D \hB \, 
\exp \!\left( -\frac{1}{\gs} 
\tr\!\left[W(\Phi) - \hW(\hPhi) - \hB \Phi B + B \hPhi \hB \right] \right)\,,
\ee
where $\Phi$ is an $M{\times}M$ matrix, 
$\hPhi$ is an $\hM{\times}\hM$ matrix, 
$B$ is an $M{\times}\hM$ matrix, 
and $\hB$ is an $\hM{\times}M$ matrix.
In the perturbative approach, the matrix integral (\ref{ZUU}) is 
evaluated about the following extremal point of the potential
\be
\label{extremalUU}
\Phi_0  = \pmatrix{ e_1 \1_{M_1}& \!\!\!\! 0& \!\!\cdots& \!\! 0 \cr
                    0& \!\!\!\! e_2 \1_{M_2}& \!\!\cdots& \!\! 0 \cr
                    \vdots& \!\!\!\! \vdots& \!\!\ddots& \!\! \vdots \cr
                    0& \!\!\!\! 0& \!\! \cdots&  \!\! e_N \1_{M_N} },\quad
\hPhi_0  = \pmatrix{ \he_1 \1_{\hM_1}& \!\!\!\! 0& \!\!\cdots&\!\! 0 \cr
                    0& \!\!\!\! \he_2 \1_{\hM_2}& \!\!\cdots& \!\!0 \cr
                    \vdots& \!\!\!\! \vdots& \!\!\ddots& \!\!\vdots \cr
                    0& \!\!\!\! 0& \!\!\cdots&  \!\!\he_N \1_{\hM_N} },\quad
B_0 =\hB_0=0,
\ee
where $\sum_i M_i =M$ and $ \sum_i \hM_i = \hM$.
This choice of vacuum breaks the 
$\U(M){\times}\U(\hM)$ symmetry to 
$G = \prod_{i=1}^N  U(M_i){ \times }\prod_{i=1}^N  U(\hM_i)$. 
Writing $\Phi = \Phi_0 + \Psi$ and  $\hPhi = \hPhi_0 + \hPsi$,   
one finds that the off-diagonal fields $\Psi_{ij}$ and $\hPsi_{ij}$
({\small $i \neq j$})
have vanishing contributions to the quadratic part of the action;  
these fields are zero modes and correspond to gauge degrees 
of freedom \cite{Dijkgraaf:2002d}.  
We fix the gauge $\Psi_{ij}=\hPsi_{ij} = 0$ ({\small $i \neq j$}) and 
introduce Grassmann-odd ghost matrices, exactly as in 
refs.~\cite{Dijkgraaf:2002d, Naculich:2002a},
to which we refer the reader for further details.
The bifundamental field $B$ 
(not to be confused with the ghost field in 
refs.~\cite{Dijkgraaf:2002d, Naculich:2002a}) 
can be written
\be                     
B = \pmatrix{ B_{11}& B_{12}& \cdots& B_{1N}  \cr
              B_{21}& B_{22}& \cdots& B_{2N}  \cr
              \vdots& \vdots& \ddots& \vdots  \cr
              B_{N1}& B_{N2} & \cdots&  B_{NN} },
\ee
where $B_{ij}$ is an $M_i \,{\times} \hM_j$ matrix,
and similarly for $\hB$.
Expanding about the vacuum (\ref{extremalUU}) one finds 
(in the $\Psi_{ij} = \hPsi_{ij} = 0$ ({\small $i \neq j$}) gauge)
\be \label{BBvertices}
\tr\!\left[\hB \Phi B - B \hPhi \hB \right] 
= \sum_{i,j} (e_i - \he_j)  \tr (\hB_{ji} B_{ij})
+ \sum_{i,j} \tr (\hB_{ji} \Psi_{ii} B_{ij} -  B_{ij} \hPsi_{jj} \hB_{ji} )\,.
\ee
Hence, the matrix integral over the quadratic action contains 
the $e_i$-dependent contribution from the bifundamental fields:
\be \label{quadB}
\prod_{i, j} \left( 1\over e_i - \he_j \right)^{M_i \hM_j}
\ee
and the trilinear pieces of (\ref{BBvertices}) contribute
$\hB \Psi B$ and $B \hPsi \hB$ vertices to the Feynman diagrams 
(in addition to the vertices considered in ref.~\cite{Naculich:2002a}).

We are interested in the planar limit of the matrix model, 
i.e.~the limit in which $\gs \rar 0$ and $M_i$, $\hM_i$ $\rar \infty$, 
keeping $\gs M_i$ and $\gs \hM_i$ fixed. 
The connected diagrams of the 
perturbative expansion of $\,\log Z$ may be organized,
using the standard double-line notation, 
in a topological expansion characterized by the 
Euler characteristic $\chi$ of the surface 
in which the diagram is embedded \cite{tHooft:1974}
\be
\log Z =  \sum_{\chi \le 2} \gs^{-\chi} F_{\chi} (S, \hS) 
\quad {\rm where~} \quad S_i \equiv \gs M_i, \quad \hS_i \equiv \gs \hM_i\,,
\ee
where $\chi = 2 {-} 2g $ with $g$ the genus.
In the planar limit, the leading contribution 
\be
\fre (S, \hS) \equiv F_{\chi=2} (S, \hS)= \gs^2 \log  Z \sphere
\ee 
comes from the connected diagrams that can be drawn on the sphere ($\chi=2$).
We will need the 
contributions to $\fre (S,\hS)$ up to cubic order in $S$ and $\hS$. 
The explicit formul\ae{} can be found in 
eqs.~(\ref{freeenergytwoUU}) and (\ref{freeenergythreeUU})
in appendix \ref{appdet}.

To relate the matrix model and its free energy
to the $\cN=2$  $\UxU$
gauge theory broken to $\prod_i \U(N_i) \times \prod_i \U(\hN_i)$, 
one introduces, following Dijkgraaf and Vafa,
\be
\label{WeffdefUU}
\Weff (S,\hS) 
= - \sum_i  N_i {\partial  \over \partial S_i} \fre (S,\hS)  
  - \sum_i  \hN_i {\partial  \over \partial \hS_i} \fre (S,\hS)  
\ee
where we have dropped terms linear in $S_i$ and $\hS_i$.
Since we are examining a generic point on the Coulomb branch
of the $\cN=2$ theory,  
which breaks $\U(N) {\times} \U(N) $ to $\U(1)^{2N}$, 
we set  $N_i=\hN_i=1$.
Next, one extremizes the effective superpotential with respect to $S_i$ 
and $\hS_i$ :
\be \label{WextremeUU}
{\partial \Weff (S, \hS) \over \partial S_i} 
\bigg|_{S_j = \vev{S_j}, \hS_j = \vev{\hS_j}}  =0\,,
\qquad
{\partial \Weff (S, \hS) \over \partial \hS_i} 
\bigg|_{S_j = \vev{S_j}, \hS_j = \vev{\hS_j} } =0\,.
\ee
The solutions for $\vev{S_i}$, $\vev{\hS_i}$ can be 
evaluated in an expansion in $\La$. 
The lowest-order contributions are 
\be
\label{SvevUU}
\vev{S_i} = 
{\alpha \Ti \over R_i}  \LU \,, \qquad \qquad
\vev{\hS_i} = 
 -  {\alpha \hTi \over \hR_i}  \LU \,,
\ee
where 
\be
\label{defTR}
\Ti = \prod_{j=1}^N (e_i - \he_j), \qquad
\hTi = \prod_{j=1}^N (\he_i - e_j), \qquad
R_i = \prod_{j \neq i} (e_i-e_j),\qquad
\hR_i = \prod_{j \neq i} (\he_i-\he_j)
\ee
and  various constants 
have been absorbed into a redefinition of the cut-off $\La$.
In sec.~\ref{sUUperttau},
we will also need the next-to-leading-order contributions; these are 
given in (\ref{SvevUU2}).

\subsubsection{Relation between $a_i$ and $e_i$}

Before computing $\tau_{ij}$ and the $\cN=2$ prepotential,
we must determine the relation between $e_i$ and 
the periods $a_i$ of the SW differential.
In ref.~\cite{Naculich:2002a}, we proposed a  
definition of $a_i$ within the context of the 
perturbation expansion of the matrix model, 
without referring to the Seiberg-Witten curve or differential.
As in refs.~\cite{Naculich:2002a, Naculich:2002b},
$a_i$ and $\ha_i$ can be determined perturbatively via 
(setting $N_i  = \hN_i = 1$)
\bea
\label{amatUU}
a_i &=&  e_i + 
 \left[ 
\sumjN {\pa \over \pa S_j} \gs \tadpole 
        + \sumjN {\pa \over \pa \hS_j} \gs \tadpole  
\right]_{\vev{S},\vev{\hS} } 
\non\\
\ha_i &=&  \he_i + 
\left[ 
\sumjN {\pa \over \pa \hS_j} \gs \htadpole  + 
\sumjN {\pa \over \pa S_j} \gs \htadpole 
\right]_{\vev{S},\vev{\hS} } 
\eea
where $\tadpole$ ($\htadpole$) is obtained by calculating 
all connected planar tadpole diagrams 
with an external $\Psi_{ii}$ ($\hPsi_{ii}$) leg 
that can be drawn on a sphere.
In addition to the tadpole diagrams 
discussed in ref.~\cite{Naculich:2002a},
there are diagrams with $B$, $\hB$ (bifundamental) loops.
The total contribution to the tadpole quadratic in $S$ and $\hS$ is 
\bea
\label{tadpoleUU}
\tadpole &=&
{1 \over \al \gs} 
\left[- \sum_{j\neq i} \frac{S_i^2 }{R_i \eij}
+ \sum_{j\neq i} 2\frac{S_i S_j}{R_i \eij} 
- \sum_{j} \frac{S_i \hS_j}{R_i \hij} 
\right] \non \\
\htadpole &=& 
-\, {1 \over \al \gs} 
\left[- \sum_{j\neq i} \frac{\hS_i^2 }{\hR_i \heij}
+ \sum_{j\neq i} 2\frac{\hS_i \hS_j}{\hR_i \heij} 
- \sum_{j} \frac{\hS_i S_j}{\hR_i \hhij} 
\right]\, 
\eea
where $\eij = e_i-e_j$, $\hij = e_i-\he_j$  and $\hhij = \he_i - e_j = -\hji$.
Inserting these results into eq.~(\ref{amatUU}),
and evaluating the resulting expression using eq.~(\ref{SvevUU}),
one finds 
\bea
\label{aeUU}
a_i &=& e_i +  \LU
\left(   {2 \over R_i} \sum_{j\neq i}\frac{\Tj }{R_j \eij} 
       + {1 \over R_i} \sum_{j} \frac{\hTj }{\hR_j \hij} 
       - {\Ti  \over R_i^2}\sum_{j}\frac{1}{\hij} 
\right)
+ \cO(\La^{2N}) \non\\
\ha_i &=& \he_i +  \LU 
\left(   {2 \over \hR_i} \sum_{j\neq i}\frac{\hTj }{\hR_j \heij} 
       + {1 \over \hR_i} \sum_{j} \frac{\Tj }{R_j \hhij} 
       - {\hTi  \over \hR_i^2}\sum_{j}\frac{1}{\hhij} 
\right)
+ \cO(\La^{2N})\,. 
\eea
By using 
\be 
\label{TRident}
\left[ {\hW'(z)\over W'(z)} - 1\right] = \sum_i {T_i \over R_i (z-e_i)}, 
\qquad
\left[ {W'(z)\over \hW'(z)} - 1\right] = \sum_i {\hT_i \over \hR_i (z-\he_i)}
\ee
one may show that
\be
\label{UUident}
\sum_{j\neq i}\frac{\Tj }{R_j \eij}  =
{\Ti\over R_i} 
\left[  
- \sum_{j\neq i} {1 \over \eij} + \sum_{j} {1 \over \hij} -1 \right],
\qquad
\sum_{j} \frac{\hTj }{\hR_j \hij} = -1
\ee
so that eq.~(\ref{aeUU}) may be rewritten\footnote{
In previous work \cite{Naculich:2002a,Naculich:2002b} 
we used similar identities at 
intermediate stages of the calculations. However, in the calculation of 
the $\cN=2$ prepotential, it is more efficient to 
work with the expressions that 
come naturally out of the matrix-model calculation. 
To compare with results obtained using M-theory at intermediate stages, 
identities generally have to be used (as we did to obtain (\ref{U2ai})).
}
\be \label{U2ai}
a_i = e_i +  \LU
\left( - {2 \Ti \over R_i^2} \sum_{j\neq i}\frac{1}{\eij} 
       + {\Ti  \over R_i^2}  \sum_{j}\frac{1}{\hij} 
       - {3 \over R_i} \right)
+ \cO(\La^{2N}) 
\ee
and similarly for $\ha_i$.
We cannot yet compare this expression with the SW result
obtained in \cite{Ennes:1998c}, because the relation between 
the roots $e_i$ of $W'(z)$ 
and the roots $e'_i$ of $P(z)$ (cf.~(\ref{U2Mcurve})) has
not yet been determined. This will be done in sec.~\ref{U2metI}.

\subsubsection{Perturbative calculation of $\tau_{ij}$}
\label{sUUperttau}

Following Dijkgraaf and Vafa, the gauge coupling matrix 
$\tau_{ij}$ is related to the planar  
free energy $F_{s}$ of the matrix model by 
\be \label{UUtauij}
\tau_{ij} 
= {1 \over 2 \pi i} \frac{\pa^2 F_{s} }{\pa S_i\pa S_j}\vevShS,
\qquad
\tau_{i \hj} 
= {1 \over 2 \pi i} \frac{\pa^2 F_{s} }{\pa S_i\pa \hS_j}\vevShS,
\qquad
\tau_{\hi \hj} 
={1 \over 2 \pi i}  \frac{\pa^2 F_{s} }{\pa \hS_i \pa \hS_j}\vevShS.
\ee 
We may calculate these expressions perturbatively using 
eqs.~(\ref{freeenergytwoUU}), (\ref{freeenergythreeUU}),
and~(\ref{SvevUU2}), and 
finally, use eq.~({\ref{aeUU}) to re-express the entire expression
in terms of $a_i$ rather than $e_i$. 
The resulting perturbative and one-instanton contributions
to the gauge coupling matrix are given in 
eqs.~(\ref{UUtaupert})--(\ref{taumixed})
in appendix \ref{appdet}.  One may verify that 
(\ref{UUtaupert})--(\ref{taumixed}) can be written as 
\be
\tau_{ij} = {\pa^2 \pree \over \pa a_i \pa a_j},
\quad
\tau_{i\hj} = {\pa^2 \pree \over \pa a_i \pa \ha_j},
\quad
\tau_{\hi\hj} = {\pa^2 \pree \over  \pa \ha_i \pa \ha_j}
\ee
with (up to a quadratic polynomial)
\bea
\label{UUpre}
2\pi i\pree
&=& 
-{\ts \frac{1}{4}}\sum_i \sum_{j\neq i} (a_i-a_j)^2 
\log \left(a_i-a_j \over \La \right)^2 
-{\ts \frac{1}{4}}\sum_i \sum_{j\neq i} (\ha_i-\ha_j)^2 
\log \left(\ha_i-\ha_j \over \La \right)^2 
\non\\
&&+{\ts \frac{1}{4}}\sum_{i,j} (a_i-\ha_j)^2 
\log \left(a_i-\ha_j \over \La \right)^2 
+  \LU  
\sum_j \left[\frac{T_j}{R_j^2} +  \frac{\hT_j}{\hR_j^2}\right]
+ \cO(\La^{2N})\,.
\eea
which agrees perfectly with (version 2 of) ref.~\cite{Ennes:1998c}.

\subsection{ {\large $\U(N)$} with $\protect\Ysymm$ or  $\protect\Yasymm$ }
\label{sPertSA}

Consider the $\cN=1$ $\U(N)$ 
supersymmetric gauge theory with one chiral superfield 
$\phi_i{}^j$  transforming in the adjoint representation 
of the gauge group, 
one chiral superfield $x_{ij}$ transforming in either the symmetric  
($\Ysymm$)
or the antisymmetric 
($\Yasymm$) 
representation, and one chiral 
superfield $\tx^{ij}$
transforming in the conjugate representation. 
We treat the cases of the symmetric and 
antisymmetric representations simultaneously  
by assuming that $x$, $\tx$ satisfy 
$x^T = \beta x$ and $\tx^T = \beta \tx$, 
where $\beta =1$ for the symmetric representation 
and $\beta =-1$ for the antisymmetric representation.
The superpotential of the gauge theory is taken to be of the 
form\footnote{A mass term $m\, \tr(\tx\, x)$ for 
the matter hypermultiplet can be 
introduced by shifting $\phi \rar \phi -m$ and $e_i \rar e_i - m$.}
\be \label{USAsuppot}
\cW(\phi,x,\tx) = \tr[W(\phi) - \tx\, \phi \,x]\,, 
\ee
where $W(\phi)$ is a polynomial such that 
$W'(\x) = \al \prod_{i=1}^N(\x-e_i)$. 
This superpotential can be viewed as a deformation of an $\cN=2$ theory,
which is recovered when $\al \to 0$,
restoring $\cN=2$ supersymmetry.  

As discussed in \cite{Klemm:2003} there are several classical ground 
states of the superpotential (\ref{USAsuppot}).
One such ground state is $\phi = e_i \1$ and $x=\tx=0$. 
Another one is $\phi =0$, $x = E$ and 
$\tx = W'(0) E^{-1}$, 
where $E=\1$ for $\Ysymm$ and $E=J$ for $\Yasymm$, 
where $J$ is the usual Sp-unit.  
There are also additional ground states as discussed 
in \cite{Klemm:2003}, but these will play no role in our discussion. 
(Similar extra vacua are also present \cite{Cachazo:2001,Hofman:2002} in the 
$\U(N){\times}\U(N)$ theory discussed above.) 
A more general vacuum is obtained by combining the above possibilities. 
In a block-diagonal basis, 
one ground state is 
$\phi = \diag(0_{N_0}, e_1 \1_{N_1} ,\ldots ,e_k \1_{N_k}$), 
with $N=N_0 + \sum_{i=1}^k N_i$
(where $N_0$ is even for $\Yasymm$)
and $x$ and $\tx$ have vanishing entries 
except for the  $N_0 {\times} N_0$ blocks
$x_{00}=E$, 
$\tx_{00} = W'(0) E^{-1}$.
Such a vacuum breaks $\U(N)$ down to \cite{Klemm:2003} 
$\SO(N_0){\times}\prod_i\U(N_i)$ for $\Ysymm$ or
$\Sp(N_0){\times}\prod_i\U(N_i)$ for $\Yasymm$.

We want to freeze the $\cN=2$ moduli to a generic,
but fixed, point on the Coulomb branch of the
$\cN=2$ theory. This is accomplished by breaking $\U(N)$ down to 
$\U(1)^N$, i.e.~choosing $N_i=1$ and $N_0=0$.

The $\U(M)$ matrix model associated with this gauge 
theory \cite{Klemm:2003,Naculich:2003}
has partition function\footnote{As in 
the previous section, we use capital letters 
to denote matrix model quantities. All matrix indices run over $M$ values.}
\be \label{ZSA}
Z ={1 \over {\rm vol~} G}
  \int \D \Phi \, \D X \, \D \tX \, 
\exp \!\left( 
-\frac{1}{\gs} \tr\!\left[W(\Phi) - \tX \Phi X \right] 
\right)\,,
\ee
where $X^T = \beta X$ and $\tX^T=\beta\tX$.
In the perturbative approach, the matrix integral (\ref{ZSA}) is 
evaluated about an extremal point of $\cW(\Phi,X,\tX)$.

Based on previous experience, it would seem natural to expand 
around a matrix-model vacuum similar to the gauge theory vacuum 
but with $N_i=1$ and $N_0=0$ replaced by 
$M_i$ and $M_0$ such that $M_i\neq 0$ and $M_0=0$. 
This will indeed turn out to be the right procedure for the 
$\U(N)+\Ysymm$ theory. 
However, as we will see, it is not the right procedure for 
the $\U(N)+\Yasymm$ theory.  
Instead, for this theory, we will take $M_0 \neq 0$ 
(in fact we will take the 
limit $M_0\rar\infty$, with $\gs M_0$ finite),
even though $N_0=0$.
We do not have an {\it a priori} reason for making this choice of vacuum. 
If one does not include the extra $M_0 \times M_0$ block for the 
matrix model corresponding to the $\U(N)+\Yasymm$ theory,
one still gets an (apparently) self-consistent result, 
but one which does not agree with the prepotential, 
SW curve, or SW differential derived from 
M-theory~\cite{Landsteiner:1998a,Naculich:1998a,Nekrasov:2002} 
(which have passed several consistency tests).
Only if one includes the extra block 
does one get a result that is in agreement with previous results.\footnote{
{\it Added in} {\bf v4}:  Although the choice of vacuum appears ad hoc
within the purely perturbative framework,
it is quite likely that the vacuum is uniquely determined
through the exact determination of the curve via Abel's
theorem (method II in section 4.2.2 of this paper), or equivalently,
through the condition that all the periods of $T(z)$
are integer-valued \cite{Cachazo:2003b}.
See refs.~\cite{Casero:2003gr} and \cite{Cachazo:2003kx}
for the use of the
latter criterion to determine the vacua in related theories.
(The $M_0 \times M_0$ block must be included whenever
an additional cut opens up in the algebraic curve around $z=0$.)}

We will decompose all matrices $\Ups$ as
\be                     
\Ups = \pmatrix{ \Ups_{00}& \Ups_{01}& \cdots& \Ups_{0N}  \cr
              \Ups_{10}& \Ups_{11}& \cdots& \Ups_{1N}  \cr
              \vdots& \vdots& \ddots& \vdots  \cr
              \Ups_{N0}& \Ups_{N1} & \cdots&  \Ups_{NN} }\,,
\ee
where $\Ups_{ij}$ is an $M_i \,{\times} M_j$ matrix, $\Ups_{i0}$ is an
$M_i{\times}M_0$ matrix, $\Ups_{00}$ is an $M_0{\times}M_0$ matrix 
(where $M_0$ is even for $\Yasymm$).
Throughout we use $i,j =1,\ldots,N$, 
displaying the index-$0$ terms explicitly. 

We evaluate the matrix integral (\ref{ZSA}) 
about the following extremal point of $W(\Phi,X,\tX)$
\be
\label{extremal1}
\Phi_0  = \pmatrix{ \,0_{M_0} & 0 & \cdots& 0  \cr
                    0& e_1 \1_{M_1}& \cdots& 0 \cr
                    \vdots& \vdots& \ddots& \vdots \cr
                    0& 0& \cdots&  e_N \1_{M_N} },\qquad
\ee
and
\be \label{extremal2}
X_0 = \pmatrix{ E & 0 & \cdots& 0  \cr
                    0 & 0 & \cdots& 0 \cr
                    \vdots& \vdots& \ddots& \vdots \cr
                    0& 0& \cdots& 0 },\qquad
\tX_0 = W'(0) \pmatrix{ E^{-1} & 0 & \cdots& 0  \cr
                    0 & 0 & \cdots& 0 \cr
                    \vdots& \vdots& \ddots& \vdots \cr
                    0& 0& \cdots&  0 },\qquad
\ee
where $M_0 + \sum_{i=1}^N M_i = M$, 
and as before $E=\1_{M_0}$ for $\Ysymm$,
and $E=J$ for $\Yasymm$, 
where $J$ is the antisymmetric $\Sp(M_0)$ unit.
This choice of vacuum breaks the 
$\U(M)$ symmetry to  \cite{Klemm:2003} 
$G = \SO(M_0) {\times} \prod_{i=1}^N  U(M_i)$ for $\Ysymm$, and 
$G = \Sp(M_0) {\times} \prod_{i=1}^N  U(M_i)$ for $\Yasymm$.

The potential in (\ref{ZSA}) is invariant under the gauge symmetry 
\be
\de \Phi = [\xi, \Phi] \,, \quad \de X = \xi X + X \xi^T \,, \quad
\de \tX = -\xi^T \tX - \tX \xi\,.
\ee 
Writing $\Phi = \Phi_0 + \Psi$ and  $X=X_0 + Y$,
we fix the gauge $\Psi_{ij} = 0$ 
({\small$i\neq j$}), $\Psi_{0i}=0$, $\Psi_{i0}=0$, 
$Y_{00} = 0$. Following \cite{Klemm:2003} we use the BRST approach and 
introduce the gauge-fixing fermion
\be
\Theta = \sum_{i \neq j} 
\tr(\bC_{ij} \Phi_{ji}) + \sum_i \tr(\bC_{i0} \Phi_{0i}) 
+ \sum_i \tr(\bC_{0i} \Phi_{i0}) + \tr(\bC_{00} [X_{00} - E]) 
\ee
where the $\bC$'s are Grassmann-odd. 
The relevant BRST transformations are \cite{Klemm:2003}
\be \label{brst}
s\, \Phi = [C,\Phi] \,, \quad s\, X = C X + X C^T \,, 
\quad s\,\bC = D 
\ee
where $C$ and $D$ are Grassmann-odd matrices. 
The gauge-fixing action $S_{\rm gf} = s\, \Theta$ 
becomes, after using (\ref{brst}) and integrating 
out the $D$'s which act as Lagrange multipliers implementing the gauge 
choice\footnote{In refs.~\cite{Dijkgraaf:2002d, Naculich:2002a} the notation 
$B$, $C$ was used for the ghost fields instead of $\bC$ and $C$.}
\bea \label{ghosts}
&&\sum_{i \neq j} \tr(\bC_{ij}[C_{ji} \Phi_{ii}-\Phi_{jj} C_{ji}]) 
+ \sum_i \tr(\bC_{i0}[C_{0i} \Phi_{ii} - \Phi_{00} C_{0i}])
+ \sum_i \tr(\bC_{0i}[C_{i0} \Phi_{00} - \Phi_{ii} C_{i0}]) \non \\
&& \,\,\, + \, \tr(\bC_{00}[C_{00} E + E C_{00}^T])
+ \sum_i\tr(\bC_{00}[C_{0i} Y_{i0} + Y_{0i} C^T_{i0}])
\eea

Expanding about the vacuum (\ref{extremal1}), (\ref{extremal2}) 
using the above gauge one finds 
(in addition to the quadratic terms in eq.~(\ref{ghosts}))
the quadratic part of the action
\bea \label{quadr}
\tr[W(\Phi) - \tX \Phi X ] 
&=&  \half \al \sumN R_i \tr(\Psi^2_{ii}) 
+\half \al R_0 \tr(\Psi_{00}^2)
- \tr(\tY_{00}\Psi_{00}E) \\
&&  - \sumN e_i \tr(\tY_{0i} Y_{i0})
- \sumN e_i \tr \left( \tY_{ii} Y_{ii}\right)
- \sum_{i<j} (e_i + e_j)  \tr (\tY_{ji} Y_{ij}) + \cdots \non
\eea
where 
\be
R_i = \prod_{j \neq i} (e_i -e_j), \qquad
R_0 = {W''(0) \over \al}  = - \prod_i(-e_i) \sum_j {1 \over e_j}\,.
\ee
{}From this we see \cite{Klemm:2003} that the antisymmetric 
matrix $\tY_{00}$ acts as a Lagrange multiplier implementing the 
constraint 
$\Psi_{00} E + \beta  (\Psi_{00} E)^T =\Psi_{00} E + E \Psi_{00}^T =0$, 
i.e.~$\Psi_{00}\in\so(M_0)$ for $\Ysymm$ and
$\Psi_{00}\in\spl(M_0)$ for $\Yasymm$.

The matrix integral over the quadratic action contains 
the $e_i$-dependent contributions
\be \label{quadX}
\left( 1 \over R_0 \right)^{\fourth M_0^2 - \fourth \beta M_0}
\prod_{i} \left[ \left( 1 \over R_i \right)^{\half M_i^2} 
 \! \left( 1\over e_i \right)^{\half M_i (M_i+\beta) }
 \! \left( e_i \right)^{M_0 M_i } \right]
\prod_{i<j} \left[ \left( e_i - e_j \right)^{2 M_i M_j} 
 \!\left( 1\over e_i + e_j \right)^{M_i M_j} \right].
\ee
The following terms in the action
(in addition to those considered in ref.~\cite{Naculich:2002a}
and the trilinear terms in eq.~(\ref{ghosts})) 
contribute cubic vertices to the Feynman diagrams:
\be
-\sumN  \left[ \tr \left( \tY_{i0} \Psi_{00}  Y_{0i}\right) 
+ \tr \left( \tY_{0i} \Psi_{ii}  Y_{i0}\right) 
+ \tr \left( \tY_{ii} \Psi_{ii}  Y_{ii}\right) \right]
- \sum_{i<j} \tr (\tY_{ji} \Psi_{ii} Y_{ij} +  \tY_{ij}  \Psi_{jj} Y_{ji})\,.
\ee
We are interested in the planar limit of the matrix model, 
i.e.~the limit in which $\gs \rar 0$ 
and $M_i, M_0 \rar \infty$, keeping 
$\gs M_i$, $\gs M_0$ fixed. 
The connected diagrams of the 
perturbative expansion of $\,\log Z$ may be organized,
using the standard double-line notation, 
in a topological expansion characterized by the Euler characteristic $\chi$ of 
the surface in which the diagram is embedded \cite{tHooft:1974}
\be
\log Z =  \sum_{\chi \le 2} \gs^{-\chi} F_{\chi} (S, \mS) 
\ee
where $\chi = 2 {-} 2g {-} q$ with $g$ the genus 
(number of handles) and $q$ the number of crosscaps. 
In the Feynman diagrams, we generally replace 
$\gs M_i$ by $S_i$ and $\gs M_0$ by $S_0$, 
but for the inner index-loop of an $X_{ij}$, $\tX_{ij}$ loop  
we write $\gs M_i = \mS_i$ since the arrow on the inner index-loop 
runs parallel
to the outer index-loop, opposite to the direction in which it would run 
for the adjoint representation
This will be important when (but not until) we calculate $\tau_{ij}$
in sec.~\ref{sSAperttau}.

In the planar limit,  
the leading contribution to the matrix integral
comes from the planar diagrams that can be drawn on the sphere ($\chi=2$),
\be
\free \equiv F_{\chi=2} (S, \mS)= \gs^2 \log  Z \sphere
\ee 
The subleading contribution comes from planar 
diagrams that can be drawn on $\R \PP^2$, which is 
a sphere with one cross-cap inserted ($\chi=1$)
\be
\frp \equiv F_{\chi=1} (S) =  \gs \log  Z \rp
\ee
To evaluate the $(1/{\rm vol~}G)$ prefactor in (\ref{ZSA}) we 
need (see e.g.~\cite{Ooguri:2002,Ashok:2002,Klemm:2003})
\bea \label{volsp}
\log {\rm vol}(\SO(M_0)) &=& 
-\fourth M_0^2 \log M_0 +\fourth M_0 \log M_0 +\ldots  \non\\
\log {\rm vol}(\Sp(M_0)) &=& 
-\fourth M_0^2 \log M_0 -\fourth M_0 \log M_0 +\ldots  \\
\log {\rm vol}(\U(M_i)) &=& 
-\half M_i^2 \log M_i +\ldots  \non
\eea
These results together with the integration over the quadratic fields 
of the matrix-model partition function yields
(up to an $e_i$-independent quadratic monomial in the $S$'s) 
\bea
\label{freeenergytwo}
\free& =& - \sumN S_i W(e_i)  
 + \half \sumN S_i^2 \log \left( S_i\over \alpha R_i \La^2 \right) 
+ \sumN \sum_{j\neq i}  S_i S_j \log\left(\eij \over \La \right) +  \non\\
&& 
-\,  \half \sumN S_i \mS_i \log \left(  e_i \over \La \right) 
- \half  \sumN \sum_{j\neq i}  S_i \mS_j \log\left(\gij \over \La \right) 
+  \sumN S_0 S_i \log \left(\frac{e_i}{\La} \right) \non \\ 
&& 
+ \, \fourth S_0^2 
\log\left(\frac{S_0}{ \al R_0 \La^2}\right)  + 
\sum_{n \ge 3} \freen
\eea
where $\eij = e_i-e_j$ and  $\gij = e_i+e_j$.
The term $\freen$ is an $n$th order polynomial in $S_i$ and $\mS_i$
arising from planar loop diagrams built from the interaction
vertices.
We will also need the cubic contribution $\freethree$; 
the explicit expression can be found 
in eq.~(\ref{freeenergythree}) in appendix \ref{appdet}. 

Next we turn to the sub-leading contributions to the free energy.
{}From eqs.~(\ref{ghosts}), (\ref{quadr}), (\ref{volsp}) one 
finds (up to an $e_i$-independent part 
linear in $S_i$)\footnote{\label{drfoot}
We note that the sphere and $\sR\sPP^2$ contributions to the free
energy obey the relation \cite{Ita:2002,Ashok:2002,Janik:2002}
\[ 
\frp  = -{\beta \over 2}{\partial  \over \partial S_0} \fre (S,S) .
\]
}
\bea
\frp & =& - \half \beta \, \sumN  S_i \log\left(\frac{e_i}{\La} \right) 
- \fourth \beta S_0 
\log\left(\frac{S_0}{\al R_0\La^2}\right) 
+ \sum_{n \ge 2} \frpn
\eea
where $\frpn$ is an $n$th order polynomial in $S_i$
arising from planar diagrams, 
built from the interaction vertices, that can be drawn on $\R \PP^2$.
We will need the quadratic contribution $\frptwo$; 
the explicit expression is given in (\ref{frpquad}).

To relate the matrix model 
to the $\cN=2$ U($N$) gauge theory broken to $\prod_i \U(N_i)$, 
one sets $\mS = S$ in the matrix-model free energy \cite{Naculich:2003}
and introduces~\cite{Ita:2002,Ashok:2002,Janik:2002,
Klemm:2003,Naculich:2003,Kraus:2003b}
\be
\label{Weffdef}
\Weff (S) 
= 
- \sum_i  N_i {\partial  \over \partial S_i} \fre (S,S) 
-  N_0 {\partial  \over \partial S_0} \fre (S,S) 
 - 4 \frp
\ee
where we have dropped terms linear in $S_i$.
Since we are examining a generic point on the Coulomb branch
of the $\cN=2$ theory,  
which breaks $\U(N)$ to $\U(1)^N$, we set $N_i=1$ and $N_0=0$.
Next, one extremizes the effective superpotential 
to obtain $\vev{S_i}$ and $\vev{S_0}$:
\be \label{Wextreme}
{\partial \Weff (S)  \over \partial S_i} 
\bigg|_{S_j = \vev{S_j}, S_0 = \vev{S_0} } =0\,, 
\qquad {\partial \Weff (S)  \over \partial S_0} 
\bigg|_{S_j = \vev{S_j}, S_0 = \vev{S_0} } 
=0\,.
\ee
The solution for $\vev{S_i}$ and $\vev{S_0}$
can be evaluated in an expansion in $\La$. 
The lowest-order contribution  is
\be
\label{Svev}
\vev{S_i} = 
{\alpha \Gi \over R_i}  \LX , \qquad
\Gi = e_i^{2\beta} \prod_{j} (e_i + e_j),
\qquad
R_i = \prod_{j\neq i} (e_i - e_j)
\ee
and
\be
\vev{S_0} = - \al (-1)^N \La^{-\beta N + 2} W''(0) 
\left( W'(0) \over 2 \right)^\beta
\ee
and constants have been absorbed into a redefinition of the cut-off $\La$.
The next-to-leading contribution,
which will be needed in sec.~\ref{sSAperttau},
is given in (\ref{Svev2}).

Now for $\Yasymm$, 
$\vev{S_0}$  and $\vev{S_i}$ are
both $\cO(\LA)$ and therefore both 
need to be included in a perturbative computation
to this order.
For $\Ysymm$, however, $\vev{S_0}$ is inversely
proportional to $\La$, which seems to indicate some 
inconsistency in perturbing about the vacuum 
(\ref{extremal1}), (\ref{extremal2}) when $\beta=1$.
Therefore, in the $\Ysymm$ case, 
we will simply expand around the vacuum with $M_0=0$ instead; 
equivalently, we will use
\be
\label{S0vev}
\vev{S_0}=2 \al\, \delta_{\beta,-1}  \La^{N+2} (-1)^N  \sum_j \frac{1}{e_j} 
\ee
in all expressions below.

\subsubsection{Relation between $a_i$ and $e_i$} \label{ASae}

Before computing $\tau_{ij}$ and the $\cN=2$ prepotential,
we must determine the relation between $a_i$ and $e_i$.
As in refs.~\cite{Naculich:2002a, Naculich:2002b},
$a_i$ can be determined perturbatively via 
(setting $N_i=1$ and $N_0=0$)
\be
\label{amat}
a_i =  e_i + \left[ \sumjN {\pa \over \pa S_j} \gs \tadpole 
+ 4 \rptadpole \right]_{\vev{S}}
\ee
where $\tadpole$ is obtained by calculating 
all connected planar tadpole diagrams 
with an external $\Psi_{ii}$ leg that can be drawn on a sphere, 
and $\rptadpole$ is obtained by computing 
all connected planar tadpole diagrams 
with an external $\Psi_{ii}$ leg that can be drawn on $\R \PP^2$.
(The factor of 4 in eq.~(\ref{amat}) arises from the corresponding factor in
$\Weff$ (\ref{Weffdef}), using the arguments in ref.~\cite{Naculich:2002a}.) 
The total contribution to the tadpole 
on the sphere quadratic in $S_i$ is
\be
\label{tadpoleSA}
\tadpole = 
{1 \over \al \gs} 
\left[- \sum_{j\neq i} \frac{S_i^2 }{R_i \eij}
+ \sum_{j\neq i} 2\frac{S_i S_j}{R_i \eij} 
- \sum_{j} \frac{S_i S_j}{R_i \gij} 
+ \frac{S_0 S_i }{R_i e_i}
\right]\,.
\ee
and contribution to the tadpole on $\R\PP^2$ linear in $S_i$ is 
\be
\label{rptadpoleSA}
\rptadpole = - {\beta\over\al} {S_i \over 2 R_i e_i}\,.
\ee
Inserting these results into eq.~(\ref{amat}),
and evaluating the resulting expression 
using eqs.~(\ref{Svev}) and (\ref{S0vev}),
one finds 
\bea
\label{aeSA}
a_i &=& e_i + \LX 
\left(   {2 \over R_i} \sum_{j\neq i}\frac{\Gj }{R_j \eij} 
       - {1 \over R_i} \sum_{j} \frac{\Gj }{R_j \gij} 
       - {\Gi  \over R_i^2}\sum_{j}\frac{1}{\gij} 
	- {2 \beta \Gi \over R_i^2 e_i}
\right) \non\\
&&
\qquad
+ \delta_{\beta,-1} \La^{N+2} \frac{2(-)^N }{R_i e_i} \sum_j \frac{1}{e_j} \,.
\eea
To compare these results with those found using SW theory,
we must consider the symmetric and antisymmetric cases
separately.

\vspace{.1in}
\noindent {\bf (i)}  $\U(N) + \Ysymm$ ($\beta=1$)
\vspace{.1in}

Consider the function 
\be \label{gdef}
g(z) = z^2 \prod_{j=1}^N \left({z + e_j \over z - e_j}\right) 
       -z^2 - 2 \si_1 z - 2 \si_1^2 
\ee
where $\si_1 = \sum_{i=1}^N e_i$,
and the last three terms remove the non-negative powers of the
Laurent expansion of $g(z)$.
The function $g(z)$ has only simple poles at $z=e_i$ and
so can be written in terms of its residues as, cf.~(\ref{Svev}),
\be \label{gdeff}
g(z) = \sum_i {G_i  \over R_i (z-e_i)} \,.
\ee
Using eqs.~(\ref{gdef}) and (\ref{gdeff}), one may show that
\bea
\label{symident}
\sum_{j\neq i}\frac{\Gj }{R_j \eij}  &=&
{\Gi\over R_i} 
\left[  
- \sum_{j\neq i} {1 \over \eij} + \sum_{j} {1 \over \gij} + {2 \over e_i}  
\right]
-e_i^2 - 2 \si_1 e_i - 2 \si_1^2 \,,\non\\
\sum_{j} \frac{\Gj }{R_j \gij}  &=&
e_i^2 - 2 \si_1 e_i + 2 \si_1^2 \,,
\eea
so that using eq.~(\ref{aeSA}) we get
\be 
\label{SAai}
a_i = e_i +  {\LS  \over R_i} \left[ -3 e_i^2 - 2 \si_1 e_i - 6 \si_1^2 
\right]
+  \LS {\Gi\over R_i^2} 
\left[ - \sum_{j\neq i} {2 \over \eij} + \sum_{j} {1 \over \gij} 
+ {2 \over e_i} \right] + \cO(\La^{2N -4}) 
\ee
After determining, in sec.~\ref{sSAmetI}, the relation between 
the roots $e_i$ of $W'(z)$ and 
the roots $e'_i$ of $P(z)$ in (\ref{USMcurve}),
we will be able to compare this result with that obtained
from SW theory.

\vspace{.1in}
\noindent {\bf (ii)}  $\U(N) + \Yasymm$ ($\beta=-1$)
\vspace{.1in}

Consider the function
\be \label{hdef}
h(z) = (-1)^N {W'(-z)\over W'(z)} 
= \prod_{j=1}^N \left({z + e_j \over z - e_j}\right)\,.
\ee
Now $h(z)/z^2$ has a double pole at $z=0$, which we may remove
by writing
\be \label{Hdef}
H(z) = {h(z) - h(0) - z h'(0) \over z^2}  = {(-1)^N\over z^2} 
\left[ {W'(-z)\over W'(z)} - 1 - 2 z \sum_{i=1}^N {1\over e_i} \right]\,.
\ee
The function $H(z)$ has only simple poles at $z=e_i$ and so may be written
\be \label{Hdeff}
H(z) = \sum_i {G_i  \over R_i (z-e_i)} \,.
\ee
Using eqs.~(\ref{Hdef}) and (\ref{Hdeff}),
one may show that
\bea
\label{antiident}
\sum_{j\neq i}\frac{\Gj }{R_j \eij}  &=&
{\Gi\over R_i} 
\left[  
- \sum_{j\neq i} {1 \over \eij} + \sum_{j} {1 \over \gij} - {2 \over e_i}  
\right]
- {(-)^N \over e_i^2 } \left[ 1+ 2 e_i \sum_j {1 \over e_j} \right] \,,
\non\\
\sum_{j} \frac{\Gj }{R_j \gij}  &=&
 {(-)^N \over e_i^2 } \left[ 1 - 2 e_i \sum_j {1 \over e_j} \right] \,,
\eea
so that using eq.~(\ref{aeSA}) we get
\be
a_i = e_i  - { 3 (-)^N \LA  \over R_i e^2_i } 
+  \LA {\Gi\over R_i^2} 
\left[ - \sum_{j\neq i} {2 \over \eij} + \sum_{j} {1 \over \gij} 
- {2 \over e_i} \right]
+ \cO(\La^{2N +4}) \,.
\ee
This equation, obtained entirely using matrix model methods,
precisely agrees (after letting $\La = -\La'$, cf.~eq.~(\ref{UAPW})),
with eq.~(4.2) in ref.~\cite{Naculich:1998a},
obtained using the Seiberg-Witten procedure,
because, as we will see in sec.~\ref{sSAmetI},
the roots $e_i$ of $W'(z)$ and 
the roots $e_i'$ of $P(z)$ in (\ref{UAMcurve}) coincide to this order in $\La$.

\subsubsection{Perturbative calculation of $\tau_{ij}$}
\label{sSAperttau}

For these models, in contrast to models 
containing only adjoint and fundamental 
representations~\cite{Dijkgraaf:2002e,Cachazo:2002b}
or bifundamental representations (sec. \ref{sPertUU}), 
the gauge coupling matrix $\tau_{ij}$ is {\em not} given by
the second derivative of the planar  
free energy $F_{s}$ of the matrix model: 
\be
\tau_{ij} 
\neq 
{1 \over 2\pi i} \frac{\pa^2 F_{s} }{\pa S_i\pa S_j} \bigg|_{S_k= \vev{S_k} } .
\ee 
Nevertheless, using a diagrammatic argument, a perturbative 
prescription for $\tau_{ij}$ can be given \cite{Naculich:2003}
\be \label{tauAS}
\tau_{ij} = {1\over 2\pi i} 
\left( {\partial \over \partial S_i } - {\partial \over \partial \mS_i } \right)
\left( {\partial \over \partial S_j } - {\partial \over \partial \mS_j } \right)
\free 
\bigg|_{S_k = \mS_k = \vev{S_k} } 
\ee
(It is unclear what, if any, the physical meaning of 
derivatives w.r.t.~$S_0$ is.)

In appendix \ref{appdet} we evaluate the expression (\ref{tauAS}) 
perturbatively 
up to one-instanton order, expressing the result in terms 
of $a_i$ rather than $e_i$. The result can be written as  
$ \tau_{ij} = {\pa^2 \pre / \pa a_i \pa a_j}$
with (up to a quadratic polynomial)
\bea
\label{SApre}
2\pi i\pre
&=& -{\ts \frac{1}{4}}\sum_i \sum_{j\neq i} (a_i-a_j)^2 
\log \left(a_i-a_j \over \La \right)^2 
  +{\ts \frac{1}{8}}\sum_i \sum_{j \neq i} (a_i+a_j)^2 
\log \left(a_i+a_j \over \La \right)^2  \\
&&  +\half (1 + \beta) \sum_i a_i^2 \log \left(a_i\over \La \right)^2 
 + \LX  \left[ \sum_i  {G_i \over R_i^2} 
-  \frac{2 \delta_{\beta,-1} }{\prod_i a_i}  \right]
+ \cO(\La^{2N-4\beta})\,. \non\
\eea
Later
(cf. eqs.~(\ref{UsymPW}, \ref{UAPW}))
we will see that $\La = -\La'$, 
where $\La'$ is the quantum scale used in 
the M-theory curves (\ref{USMcurve},\ref{UAMcurve}).
Taking this into account, eq.~(\ref{SApre})
precisely agrees with the calculations of the 
prepotential in refs.~\cite{Naculich:1998a,Ennes:1998b}
which utilize the SW curves for these theories derived from M-theory
\cite{Landsteiner:1998a}.

\setcounter{equation}{0}
\section{Cubic matrix-model curves}
\label{sMMcurve}

In this section we study the 
algebraic curves that arise from the planar solution
of the matrix models and show how to obtain from these the SW curves 
of the $\cN=2$ gauge theories discussed in sec.~\ref{sMtheory}. 

\subsection{{\large $\U(M){\times}U(\hM)$} matrix model 
with bifundamental matter}

The large-$M$ planar solution 
of the $\U(M){\times}\U(\hM)$ matrix model 
described in sec.~\ref{sPertUU}
was discussed in \cite{Lazaroiu:2003,Naculich:2003} 
(several of the results can also be 
found in~\cite{Kostov:1992,Dijkgraaf:2002b,Hofman:2002}).
In this approach, one defines the resolvents 
\be \label{ress}
\om(z) = \gs \bvev{ \!\tr\!\left(\frac{1}{z-\Phi}\right) \!} 
= \gs \sum_{n=0}^{\infty} z^{-n-1} \vev{ \tr\, \Phi^n}
\,, \quad 
\home(z) = \gs \bvev{ \!\tr\!\left(\frac{1}{z-\hPhi}\right) \!} 
= \gs \sum_{n=0}^{\infty} z^{-n-1} \vev{ \tr\, \hPhi^n}
\ee
where matrix-model expectation values are defined via
\be \label{U2MMexp}
\bvev{\cO(\Phi,\hPhi,B,\hB)} 
\equiv
\frac{1}{Z} \int \D \Phi \, \D \hPhi \,  \D B \, \D \hB \, 
\cO(\Phi,\hPhi,B,\hB) \, 
\e^{  -\frac{1}{\gs}
\tr\left[W(\Phi) - \hW(\hPhi) - \hB \Phi B + B \hPhi \hB \right] }\,.
\ee
It may be shown that 
\be \label{U2uis}
u_1(z) = - \om(z) + W'(z) \,, \qquad 
u_2(z) = \om(z) - \home(z) \,, \qquad 
u_3(z) = \home(z) + \hW'(z) 
\ee
(where $\om(z)$ is the leading (sphere) part of the resolvent)  
are the values of a variable $u$
on the three sheets of a Riemann surface
defined by\footnote{This equation may be obtained 
e.g.~from eq.~(2.22) in ref.~\cite{Naculich:2003} by redefining 
$u \rar -u + \third (W'(z)+\hW'(z))$
and setting $t_1(z) = s_1(z) + \third (W'(z)+\hW'(z)) r_1(z)$}
\be \label{UUMM}
(u - W'(z))\,u\,(u-\hW'(z)) =  r_1(z)\, u - t_1(z)\,.
\ee
The coefficients of the cubic curve (\ref{UUMM})
are given by \cite{Lazaroiu:2003,Naculich:2003}
\bea \label{U2r1}
r_1(z) &=&  - \gs \bvev{\tr\!\left(\frac{W'(z)-W'(\Phi)}{z-\Phi}\right) } 
+ \gs \bvev{\tr\!\left(\frac{\hW'(z)-\hW'(\hPhi)}{z-\hPhi}\right) } 
\eea
and 
\bea  \label{U2t1}
t_1(z) &=& 
-\gs\,\hW'(z) \bvev{\tr\!\left(\frac{W'(z)-W'(\Phi)}{z-\Phi}\right)}
+ \gs \,W'(z) \bvev{\tr\!\left(\frac{\hW'(z)-\hW'(\hPhi)}{z-\hPhi}\right) }
\non\\
&& - \,\gs^2 \bvev { \tr\!\left[ \frac{\D}{\D \Phi}
\left(\frac{W'(z) - W'(\Phi)}{z-\Phi}\right) \right] } 
  - \gs^2 \bvev {\tr\!\left[\frac{\D}{\D \hPhi} 
\left(\frac{\hW'(z) - \hW'(\hPhi)}{z-\hPhi}\right)\right] }  \non\\
&&
+ \,\gs \bvev{\tr\!\left(\frac{W'(z)-W'(\Phi)}{z-\Phi} W'(\Phi)\right) }
- \gs \bvev{\tr\!\left(\frac{\hW'(z)-\hW'(\hPhi)}{z-\hPhi} \hW'(\hPhi)
\right) }  
\eea
from which one can see, using eq.~(\ref{Wdef}),
that $r_1(z)$ and $t_1(z)$ are polynomials
of degree at most $N{-}1$ and $2N{-}1$, respectively, 
whose coefficients depend on the vevs 
$\vev{\tr(\Phi^k)}$ and $\vev{\tr(\hPhi^k)}$ 
with $k\le 2N{-}1$. 
At this point these vevs, and therefore 
$r_1(z)$ and $t_1(z)$, are undetermined.
We would now like to connect the above general curve 
to the cubic Seiberg-Witten curve (\ref{U2ucurve}) 
for the $\cN=2$ theory. 
We will discuss two methods.

\subsubsection{Method I: perturbative determination of the curve} 
\label{U2metI}

The planar solution to the matrix model yields a curve
(\ref{UUMM})
dependent on arbitrary polynomials $r_1(z)$ and $t_1(z)$.
As we saw above, the coefficients of these polynomials 
depend on $\vev{\tr(\Phi^k)}$ and $\vev{\tr(\hPhi^k)}$, 
which at this stage are arbitrary.
An additional condition is necessary to fix these polynomials,
namely, the extremization of $\Weff$. 
This will determine $\vev{\tr(\Phi^k)}$ and $\vev{\tr(\hPhi^k)}$, 
and thus lead to a specific form of the cubic curve.
Only then can the matrix-model curve be compared with the
Seiberg-Witten curve obtained from M-theory 
(see section \ref{sMtheory}).

One method of using the extremization of $\Weff$ 
to determine the curve employs Abel's theorem,
and was described in section 7 of ref.~\cite{Naculich:2002b} 
(see also~\cite{Cachazo:2002a,Ahn:2003a}).
This approach will be discussed below in subsection \ref{U2metII}.

However, the method using Abel's theorem is difficult to 
apply in some cases, so in this section we will present
an alternative approach that is more straightforward to implement.
This method is to evaluate $ \vev{\tr (\Phi^n)}$ perturbatively
in powers of $\Lambda$, and use the result to determine 
$r_1(z)$ and $t_1(z)$, and therefore the form of the cubic curve,
order-by-order in perturbation theory.
Although this method does not yield the exact form of the curve, 
it is a quick and efficient way of determining the 
form of the curve to lowest order in $\La$.

Expanding $\Phi = \Phi_0 + \Psi$,  
where $\Phi_0$ is given by (\ref{extremalUU})
one easily sees that 
to lowest order in perturbation theory
the matrix model expectation values $\gs \vev{\tr(\Phi^n)}$ 
are given by $\sum_i \vev{ S_i } \, e_i^n $.
Thus, writing
\be
\label{cndef}
\frac{W'(z)-W'(\Phi)}{z-\Phi} = \sum_{n=0}^{N-1} c_n(z) \Phi^n
\ee
and using $W'(e_i)=0$, we have 
\be \label{U2result}
\gs \bvev{\tr\!\left(\frac{W'(z)-W'(\Phi)}{z-\Phi}\right) } 
= \sum_n c_n(z)  \gs \vev{\tr(\Phi^n)} 
= \sum_i \vev{S_i} \sum_n c_n(z) e_i^n 
= W'(z) \sum_i {\vev{S_i} \over  z-e_i} 
\ee
where the second equality only holds to lowest order.
If one is only interested in the lowest-order contribution,
one can drop the last four terms in $t_1(z)$ (\ref{U2t1}). 
(The terms on the second line of (\ref{U2t1}) are double-trace terms;
since they contain products of at least two $S$'s,
they are at least second order in $\LU$. 
The terms on the last line of (\ref{U2t1}) vanish to lowest order 
since $W'(e_i)=\hW'(\he_i)=0$.)
Using (\ref{U2result}), one finds (to lowest order)
\bea\label{r1t1}
r_1(z)&=&  - W'(z)  \sum_i {\vev{S_i} \over  z-e_i}
+ \hW'(z) \sum_i {\vev{\hS_i} \over  z-\he_i} \,,  \non\\
t_1(z) &=& -\hW'(z) W'(z) \sum_i {\vev{S_i} \over  z-e_i}  
+ W'(z) \hW'(z) \sum_i {\vev{\hS_i} \over  z-\he_i} \,.  
\eea
Using eq.~(\ref{SvevUU}) we find,
again to lowest order,
\be\label{sum}
 \sum_i {\vev{S_i} \over  z-e_i} 
= \al \LU \sum_i {T_i \over R_i (z-e_i)}\,, \qquad\qquad
 \sum_i {\vev{\hS_i} \over  z-\he_i} 
= -\al \LU \sum_i {\hT_i \over \hR_i (z-\he_i)}\,.
\ee
Inserting eq.~(\ref{sum}) into eq.~(\ref{r1t1}) 
and using eq.~(\ref{TRident}), one obtains
\be
\label{UUrt}
r_1(z) = 0 + \cO(\Lambda^{2N} ), \qquad 
t_1(z)= - \al \LU  \left[ W'(z) - \hW'(z) \right]^2 + \cO(\La^{2N})
\ee
hence the matrix model curve is (to first order in $\LU$)
\be
\label{U2mmcurve}
(u - W'(z))\,u\,(u-\hW'(z)) = \al \LU  \left[ W'(z) - \hW'(z) \right]^2 
+ \cO(\La^{2N})
\ee
This curve is identical to the (transformed)
M-theory curve (\ref{U2ucurve}) 
provided that 
\bea \label{UUPW}
W'(z)& =& \al \left( P(z) - 3 \LUp  \right),  
\qquad u = \al \, u'
\non\\
\hW'(z) &=& \al \left( \hP(\x) - 3 \LUp  \right), 
\qquad \La = \Lap
\eea
In summary, the matrix model curve (\ref{UUMM}), 
together with the extremization of $\Weff$, 
which gives (\ref{SvevUU}) and therefore (\ref{UUrt}),
leads to the $\UxU$ SW curve (\ref{U2ucurve}).

The relation (\ref{UUPW}) 
implies that the roots
of the polynomial $P(z)  = \prod_{i=1}^N (z-e'_i)$ in the SW curve 
(\ref{U2Mcurve})
and the roots $e_i$ of the derivative of the matrix model potential 
$W'(z) = \al \prod_{i=1}^N (z-e_i)$ 
are equivalent classically, 
but differ by
\be
\label{eeprimeUU}
e'_i = e_i - {3 \LU \over R_i} + \cO(\La^{2N})
\ee
to first order in $\LU$,
and analogously for $\he_i$ and $\he'_i$.
This just amounts to a redefinition of the moduli $e_i$ and $\he_i$.
In sec.~\ref{sPertUU}
we determined the relation (\ref{U2ai})
between the SW periods $a_i$ and $e_i$.  
Combining (\ref{U2ai}) and (\ref{eeprimeUU})
allows us to write
\be \label{U2ai2}
a_i= e'_i +  \LUp {\Ti \over R_i^2} 
\left( -  \sum_{j\neq i}\frac{1}{\eij} +  \sum_{j}\frac{1}{\hij} \right)
+ \cO(\Lap\,^{2N}) 
\ee
(and a similar relation between $\ha_i$ and $\he'_i$).
This results precisely agrees with eq.~(17) of ref.~\cite{Ennes:1998c},
obtained using the Seiberg-Witten procedure.

The fact that the first-order curve (\ref{U2mmcurve})
precisely agrees with the M-theory curve (\ref{U2ucurve})
(which is believed to be the exact answer)  
points to the existence of 
a non-renormalization theorem in the matrix model
(which we have not proven). 
As shown in appendix \ref{aUNf}, a similar result holds for 
$\U(N)$ with $N_f$ fundamentals when $N_f<N$ (but not when $N_f\ge N$). 
Thus in some cases (but not always) the perturbative method 
described above actually gives exact results.

\subsubsection{Method II: exact determination of the curve 
via Abel's theorem} \label{U2metII}

In this section, we will follow the strategy of \cite{Naculich:2002b}
and discuss, using the saddle-point solution, the condition on the 
matrix-model curve imposed by extremizing $\Weff$.

The cubic curve (\ref{UUMM})  with the right-hand
side set to zero is a singular curve with singularities
at $z=e_i$ (the roots of $W'(z)$),
at $z=\he_i$ (the roots of $\hW'(z)$),
and at the roots of $W'(z)-\hW'(z)$.
Turning on the right-hand side {\it generically}
deforms the curve into a three-sheeted Riemann surface 
with (square-root) branch cuts between sheets one and two 
located near $e_i$, 
branch cuts between sheets two and three
located near $\he_i$, 
and branch cuts between sheets one and three
located near the roots of $W'(z)-\hW'(z)$
(see, e.g., \cite{Hofman:2002} for a picture of the 
cut structure of this curve).
This generic Riemann surface has genus $3N-2$.
If, however, the last described set of cuts does not open up,
the curve remains singular, having (geometric) genus $2N-2$.
(This is in fact the case for the SW curve (\ref{U2ucurve}),
agreeing with the fact that the $\cN=2$ moduli
space is $2(N-1)$-dimensional.)

To impose the extremization of $\Weff$ on the matrix model curve
(\ref{UUMM}),
we begin by expressing the leading (sphere) part 
of the free-energy of the matrix model
in an eigenvalue basis as (cf. \cite{Hofman:2002})
\bea \label{U2F0}
\fre &=&   \int \! \D \la \, \D\la' \, [ \rho(\la)\rho(\la') \log(\la-\la')
 + \hrho(\la)\hrho(\la') \log(\la-\la') - \rho(\la)\hrho(\la')\log(\la-\la') ] 
\non \\
&& \!\!\!\!\! - \int \! \D\la \, [\rho(\la) W(\la) - \hrho(\la) \hW(\la) ] 
\eea
where $\rho(\la)$ and $\hrho(\la)$ are the densities of eigenvalues
\be
\rho(\lam)= \gs \sum_i \delta(\lam-\lam_i) \,, \qquad\qquad
\hrho(\lam)= \gs \sum_i \delta(\lam-\hla_i)
\ee
(normalized as $\int \D \la\, \rho(\la) = \gs M = S$ and 
$\int \D\la \, \hrho(\la) = \gs \hM = \hS$), 
which are related to the resolvents (\ref{ress}) via 
\be
\label{UUresolvent}
\om(\x) = \int \! \D \lam {\rho(\lam) \over z - \lam } \,, \qquad \qquad
\home(\x) = \int \! \D \lam {\hrho(\lam) \over z -\lam} \,.
\ee
Next we define\footnote{In general there are also 
corresponding $S$'s for the cuts 
connecting sheets one and three (see \cite{Hofman:2002} for a discussion). 
However, these will not affect our discussion so we will suppress them. 
} 
\be
\label{Sdefinition}
S_i = -\,{1\over 2 \pi i} \oint_{A_i} \! u(z) \, \D z,
\qquad
\hS_i = {1\over 2\pi i} \oint_{\hA_i} \! u(z) \, \D z
\ee
where $A_i$ and $\hA_i$ denote contours 
around the branch cuts near $e_i$ and $\he_i$ 
on sheets one and three respectively.
Using (\ref{U2uis}) and (\ref{UUresolvent})
one can show that $S_i$ and $\hS_i$ are the 
integrated densities of eigenvalues 
along the cuts near $e_i$ and $\he_i$.
(Thus the definition (\ref{Sdefinition}) is consistent with
the perturbative definition $S_i = \gs M_i$, $\hS_i = \gs \hM_i$.)
Variations in $S_i$ and $\hS_i$ can be implemented
\cite{Ferrari:2002,Naculich:2002b}
by varying the densities
$\delta \rho (\lam) = \delta S_i \, \delta(\lam-e_i)$
and 
$\delta \hrho (\lam) = \delta \hS_i \, \delta(\lam-\he_i)$,
with $e_i$ and $\he_i$ denoting any point along the branch cuts
(cf.~\cite{Cachazo:2003b} for an alternative approach).
Specifically,
(up to terms which will not affect our discussion;
see \cite{Cachazo:2003b} for a discussion of such terms)
\bea \label{result1}
{\partial \fre \over \partial S_i}
&=&
 -  W(e_i) 
 +  \int \! \D \la [ 2 \rho(\la) \log(\la-e_i) - \hrho(\la)\log(\la-e_i) ] 
\non \\
&=&
  \int_{e_i}^I \D z  W'(z) 
 +\int \! \D \la \left[ - 2 \rho(\la) \int_{e_i}^I \D z  {1 \over z-\la}  
                   + \hrho(\la)  \int_{e_i}^I \D z  {1 \over z-\la}   \right]
\non \\
&=&
  \int_{e_i}^I \D z \left[ W'(z) - 2 \om(z) + \home(z) \right]
  = \int_{e_i}^I \D z \left[ u_1(z) -u_2(z) \right]  
  = \int_{I_2}^{I_1} u
\eea
where we have used (\ref{U2uis}), and
the last expression is interpreted as the integral of $u$ from 
$I_2$, infinity on the second sheet, to
$I_1$, infinity on the first sheet, along
a contour that passes through the cut near $e_i$.
Similarly,
\be \label{result2}
{\partial \fre \over \partial \hS_i} 
= \int_{I_3}^{I_2}  u
\ee
where the integral is taken along a contour that passes through 
the cut near $\he_i$. 
The results (\ref{result1}), (\ref{result2}) were also obtained 
in~\cite{Hofman:2002}. 

Next, we wish to extremize the effective superpotential  (\ref{WeffdefUU})
\be
\Weff 
= - \sumN  {\partial  \fre \over \partial S_i} 
  - \sumN  {\partial \fre  \over \partial \hS_i} 
\ee
(setting $N_i=\hN_i=1$) 
for the $\UxU$ theory broken down to $\U(1)^{2N}$.
This is accomplished by taking derivatives 
of $W_{\rm eff}$ w.r.t.~the $S_i$'s and $\hS_i$'s. 
However, in analogy with \cite{Cachazo:2002a,Naculich:2002b}, 
one may change variables and instead 
vary w.r.t.~the coefficients of the arbitrary polynomials 
$r_1(z)$ and $t_1(z)$ in the matrix model curve (\ref{UUMM}).
{}From (\ref{UUMM}), one may check that the derivatives
of $u$ w.r.t. the coefficients of $r_1(z)$ and $t_1(z)$ are 
holomorphic differentials on the Riemann surface\footnote{
Actually not all these differentials are holomorphic; we only 
consider the holomorphic ones.}.
We can change basis to the canonical basis of holomorphic 
differentials, $ \zeta_k$, dual to the homology basis, so that 
the conditions arising from extremizing $\Weff$ may be written as
(see ref.~\cite{Naculich:2002b} for further details)
\be
\label{Abel}
N \int_{I_3}^{I_1} \zeta_k = 0 \qquad {\rm (modulo\; period \;lattice)}\,.
\ee
This condition implies, by Abel's theorem, 
the existence of a function with an $N$th order pole at $I_1$,
an $N$th order zero at $I_3$, and regular everywhere else. 
For a generic choice of $r_1(z)$ and $t_1(z)$ such a function will not exist 
(by the Weierstrass gap theorem). Thus only in very special circumstances 
can such a function exist. 

Let us first show that the problem has a solution. 
Consider eq.~(\ref{UUMM}), with $r_1(z) = 0$ 
and\footnote{If one starts with an arbitrary $t_1(z)$ instead, 
the requirement that the function in eq.~(\ref{U2f}) 
below should have the desired 
properties implies that $t_1(z)$ has to be of this form.  
} 
$t_1 (z) \propto  (W'(z)-\hW'(z)^2$, 
i.e. a curve of the form:
\be
\label{repeat}
(u - W'(z))\,u\,(u-\hW'(z)) =  {\rm~const} \times (W'(z)-\hW'(z))^2 \,.
\ee
One can infer the asymptotic behavior of $u$ at $I_i$, infinity 
on each of the three sheets 
\be \label{asymp}
I_1: \quad u = W'(z) + \cO(z^{-1}) \,, \qquad 
I_2: \quad u = \cO(z^{-1}) \,, \qquad
I_3: \quad u = \hW'(z) + \cO(z^{-1})
\ee
(Alternatively, this can be deduced from (\ref{U2uis}) 
together with the asymptotic behavior of the resolvents.)

Now consider the function, defined on the curve (\ref{repeat}),
\be \label{U2f}
f(z) = \frac{u-\hW'(z)} { u-W'(z) }
\ee
This function has the right asymptotic properties to satisfy
Abel's theorem, but potentially has poles and zeros at finite values
$z_0$.
The denominator vanishes at any point $z_0$ at which 
$u(z_0)= W'(z_0)$ on one of the sheets. 
However, using (\ref{repeat}), 
one sees that the function is actually regular at these (singular) points. 
Similarly the potential zeros at the points where 
$u(z_0)= \hW'(z_0)$ are absent. 
Thus the function (\ref{U2f}) defined 
on the curve (\ref{repeat}) has the divisor 
implied by (\ref{Abel}) via Abel's theorem.

Assuming that the solution is unique, we find that the matrix model 
implies a Seiberg-Witten curve of the form (\ref{repeat})
(with no terms higher order in $\La$). 
This precisely agrees with the M-theory result (\ref{U2ucurve}) 
after the redefinitions (\ref{UUPW}).
Note that after these redefinitions, the function (\ref{U2f}) 
is proportional to the variable $y$ (\ref{U2y}) 
appearing in the M-theory curve (\ref{U2Mcurve}). 
We have not been able 
to show uniqueness.

\subsection{{\large $\U(M)$} matrix model with symmetric or antisymmetric matter}

The large-$M$ planar solution of the 
$\U(M)$ matrix models 
described in sec.~\ref{sPertSA}, 
was discussed in refs.~\cite{Klemm:2003,Naculich:2003} 
(these models are a slight modification of the $O(1)$ model 
described in refs.~\cite{Kostov:1989}). 
In this approach, one defines the resolvent
\be \label{res}
\om(z) = \gs \bvev{ \tr\left(\frac{1}{z-\Phi}\right) } = 
\gs \sum_{n=0}^\infty z^{-n-1}\vev{\tr(\Phi^n)}
\ee
where matrix-model expectation values are defined via
\be \label{USAMMexp}
\vev{ \cO(\Phi,X,\tX) } = 
\frac{1}{Z} \int \D \Phi \, \D X \, \D \tX \,\cO(\Phi,X,\tX)  \,
\e^{ -\frac{1}{\gs}\tr[W(\Phi) - \tX \Phi X] }\,.
\ee
Then it may be shown that
\be \label{USuis}
u_1(z) = - \om_0(z) + W'(z) \,, \quad 
u_2(z) = \om_0(z) + \om_0(-z) \,, \quad 
u_3(z) = -\om_0(-z) + W'(-z) ,
\ee
where $\om_0(z)$ is the leading (sphere) 
part of the resolvent,  
are the values of a variable $u$
on the three sheets of a Riemann surface
defined by\footnote{This may be obtained, e.g., from eq.~(3.20) 
in ref.~\cite{Naculich:2003}, by redefining
$u \rar -u + \third (W'(z)+W'(-z) )$
and setting $t_1(z) = s_1(z)  + \third (W'(z)+W'(-z)) r_1(z)$.}
\be \label{USMMcurvers}
(u - W'(z))\,u\,(u-W'(-z)) =  r_1 (z)\,u - t_1(z)
\ee
The coefficients of the cubic curve (\ref{USMMcurvers})
are given by \cite{Klemm:2003,Naculich:2003}
\bea \label{USr1}
r_1(z) &=&  - \gs \bvev{\tr\!\left(\frac{W'(z)-W'(\Phi)}{z-\Phi}\right) } 
- \gs \bvev{\tr\!\left(\frac{W'(-z)-W'(\Phi)}{-z-\Phi}\right) }  
\eea
and 
\bea  \label{USs1}
t_1(z) &=& 
\bigg[ -\gs W'(-z) \bvev{\tr\!\left(\frac{W'(z)-W'(\Phi)}{z-\Phi}\right)}
 - \gs^2 \bvev { \tr\!\left[ \frac{\D}{\D \Phi} 
\left(\frac{W'(z) - W'(\Phi)}{z-\Phi}\right) \right] } \non \\
&&\;\;\, +\,\gs \bvev{\tr\!\left(\frac{W'(z)-W'(\Phi)}{z-\Phi} W'(\Phi)
\right) }\bigg]
+ (z\leftrightarrow -z) 
\eea
from which one sees that $r_1(z)$ and $t_1(z)$ are even polynomials
of degree at most $N{-}1$ and $2N{-}2$, respectively, 
whose coefficients depend on the vevs $\vev{\tr(\Phi^k)}$ with $k\le 2N{-}1$.

\subsubsection{Method I: perturbative determination of the curve}
\label{sSAmetI}

As in section (\ref{U2metI}), we may evaluate
the polynomials $r_1$ and $t_1$ perturbatively
in $\La$, and use the result to determine the curve order-by-order 
in perturbation theory. 
Expanding $\Phi = \Phi_0 + \Psi$,  
where $\Phi_0$ is given by (\ref{extremal1}),
one sees that,
to lowest order in perturbation theory,
the matrix model expectation value $\gs \vev{\tr(\Phi^n)}$ 
is given by 
$\sum_i \vev{ S_i } \, e_i^n  + \vev{S_0} \delta_{n,0}$.
Hence, similar to eq.~(\ref{U2result}), we find
\bea\label{USAr1t1}
r_1(z)&=&-W'(z)  \sum_i {\vev{S_i} \over  z-e_i} 
- W'(z)\frac{\vev{S_0}}{z} \quad + \quad (z \to -z)   \,.
\non\\
t_1(z) &=&-W'(-z) W'(z) \sum_i {\vev{S_i} \over  z-e_i} 
- W'(-z) W'(z)\frac{\vev{S_0}}{z} \; +\;  (z \to -z) \,.
\eea
Using the lowest-order perturbative
results (\ref{Svev}), (\ref{S0vev}) we have
\be
\sum_i {\vev{S_i} \over z-e_i} 
= \al \La^{N-2\beta} \sum_i {G_i \over R_i (z-e_i)} \,, \qquad 
\qquad \vev{S_0} =  2 \al\, \delta_{\beta,-1}  \La^{N+2} 
(-1)^N  \sum_j \frac{1}{e_j} \,.
\ee
To evaluate
$ \sum_i [ G_i /R_i (z-e_i)]$,
we must consider the $\U(N) + \Ysymm$ and $\U(N) + \Yasymm$ 
cases separately.

\vspace{.1in}
\noindent {\bf (i)}  $\U(N) + \Ysymm$ ($\beta=1$)
\vspace{.1in}

Using eqs.~(\ref{gdef}) and (\ref{gdeff}), 
we find (to lowest order)
\bea
r_1(z)&=&  - \al \LS  W'(z) g(z) \quad +\quad  (z \to -z)   \\
&=& \al \LS 
\left(z^2 - (-1)^N z^2 + 2 \si_1^2 \right) \left[W'(z) + W'(-z)\right]
+ 2 \al \LS \si_1 z \left[W'(z) - W'(-z)\right]  
\non
\eea
and (also to lowest order)
\bea
t_1(z) &=& - \al \LS W'(-z) W'(z) g(z) + (z \to -z)\\
&=& - \al \LS (-1)^N z^2 \left[ W'(z) {-} W'(-z) \right]^2
+ 2 \al \LS \left(z^2{-}(-1)^N z^2 {+} 2 \si_1^2\right) W'(z) W'(-z)
\non
\eea
The bottom lines of these equations make manifest that 
$r_1(z)$ and $t_1(z)$ are (even) polynomials, 
and the top lines show that they are of degree
$N-1$ and $2N-2$ respectively.

The cubic equation (\ref{USMMcurvers}), 
with $r_1(z)$ and $t_1(z)$ as above,
may be considerably simplified by defining
\bea \label{UsymPW}
W'(z) &=& \al \left[ 
P(z) + \LS \left( -3z^2 - 2 \si_1 z - 6 \si_1^2 \right) \right] 
+\cO(\La^{2N+4})\,,
\\
u    &=&  \al \left[ u' -   2 \LS (z^2 - (-1)^N z^2 + 2 \si_1^2)\right] 
+\cO(\La^{2N+4})\,,
\non \\
\La &=& -\La' \,. \non
\eea
Then
\be
(u' - P(z) + 3 \La'^{N-2} \x^2)\,u'\,(u' - \Qx + 3 \La'^{N-2} z^2) 
 =  \La'^{N-2} z^2 (P(\x)-\Qx)^2 \non\\
+ \cO(\La'\,^{2N-4})
\ee
in agreement with the transformed M-theory curve (\ref{USucurve}).

The relation (\ref{UsymPW}) implies that the roots
of the polynomial $P(z)  = \prod_{i=1}^N (x-e'_i)$ in the SW curve 
(\ref{USMcurve})
and those of the derivative of the matrix-model potential 
$W'(z) = \prod_{i=1}^N (x-e_i)$ are equivalent classically, 
and are related by 
\be \label{eprimesym}
e_i' = e_i + {\LS \over R_i} \left( -3 e_i^2 - 2 \si_1 e_i - 6 \si_1^2 \right)
+ \cO(\La^{2N-4})
\ee
at the one-instanton level.
Combining this result with the relation (\ref{SAai})
between $a_i$ and $e_i$, we find
\be
a_i = e'_i +  
(-)^N \La'^{N-2}
  {\Gi\over R_i^2} \left[ - \sum_{j\neq i} {2 \over \eij} 
+ \sum_{j} {1 \over \gij} + {2 \over e_i} \right]
+ \cO(\La'\,^{2N -4})\,.
\ee
This equation precisely agrees with eq.~(26) of ref.~\cite{Ennes:1998b},
obtained using the Seiberg-Witten procedure.

\vspace{.1in}
\noindent {\bf (ii)}  $\U(N) + \Yasymm$ ($\beta=-1$)
\vspace{.1in}

Using eqs.~(\ref{Hdef}) and (\ref{Hdeff}),
one finds (to lowest order)
\bea
r_1(z)&=&  - \al \LA  W'(z) \left [H(z)  
+ 2 (-1)^N  \frac{1}{z} \sum_i \frac{1}{e_i} \right] 
\quad+\quad (z \to -z)   \non\\
&=& 0
\eea
and
\bea
t_1(z) &=& 
- \al \LA W'(-z) W'(z) \left [H(z)  
+ 2 (-1)^N  \frac{1}{z} \sum_i \frac{1}{e_i} \right]  
+ (z \to -z)   \non\\ 
&=& - \al (-1)^N \LA \left[ W'(z) - W'(-z) \over z \right]^2\,.
\eea
Defining
\be\label{UAPW} 
W'(z) = \al  P(z) \,, \qquad\quad
u = \al \,u' \,, \qquad\quad
\La = -\La' \,,
\ee
equation (\ref{USMMcurvers}) may be rewritten as
\be
(u' - P(z))\,u'\,(u' - \Qx ) 
 =  {\La'^{N+2} \over z^2} (P(\x)-\Qx)^2 + \cO(\La'\,^{2N+4})
\ee
which is in agreement with the (transformed) M-theory curve 
(\ref{UAucurve}).

Note that (\ref{UAPW}) implies that the roots 
of the polynomial $P(z) = \prod_{i=1}^N (x-e'_i)$ 
in the SW curve and those of the 
derivative of the matrix model potential $W'(z)$ 
coincide to the order we have calculated, 
i.e.~$e_i'=e_i + \cO(\La^{2N+4})$.  
This result was used in sec.~\ref{sPertSA} to compare the relation 
between $a_i$ and $e_i$ obtained from the matrix model with
the one derived in ref.~\cite{Naculich:1998a} using
the M-theory curve (\ref{UAMcurve}). 
The two results agree (see sec.~\ref{sPertSA}).

\subsubsection{Method II: exact determination of the curve 
via Abel's theorem}

The condition imposed on the matrix-model curve by extremizing $\Weff$ 
can also be discussed using the saddle-point solution, as
in \cite{Naculich:2002b} and in sec.~\ref{U2metII}.

The Riemann sheet structure of the generic cubic 
curve (\ref{USMMcurvers}) is similar to that for the 
$\UxU$ model discussed in subsection \ref{U2metII},
except that we are interested in the quotient of this 
curve by the involution $z\to -z$.

In an eigenvalue basis the leading (sphere) part of the free-energy 
can be written 
\be \label{USAF0}
\fre 
= - \int \! \D\la \, \rho(\la) W(\la) 
+ \int \! \D \la \, \D\la' \, [ \rho(\la)\rho(\la') \log(\la-\la')
- \half \rho(\la)\rho(\la')\log(\la+\la') ] 
\ee
and the subleading ($\R\PP^2$) part is
\be
 \frpp = -\half \beta \int \D\la \, \rho(\la) \log(\la) 
\ee
where 
\be
\rho(\lam)= \gs \sum_i \delta(\lam-\lam_i) \,,
\qquad
\om_0(\x) = \int \! \D \lam {\rho(\lam) \over z - \lam } \,. 
\ee
We define\footnote{
As in the $\UxU$ theory, there are in general additional $S$'s 
corresponding to the other (possible) cuts of the curve. 
In particular, there is a variable $S_0$ corresponding 
to the (possible) cut between sheets one and three around 
the point $z=0$
\[
S_0 = -\frac{1}{2\pi i}\oint_{0_1} \! u(z)\, \D z 
\]
} 
\be
\label{SSAdefinition}
S_i = -\,{1\over 2 \pi i} \oint_{A_i} \! u(z) \, \D z,
\ee
where $A_i$ denote contours 
around the branch cuts near $e_i$ on sheet one.
As before one can show that 
$S_i$ is the integrated density of eigenvalues 
along the cut near $e_i$
(so (\ref{SSAdefinition}) is consistent with 
the perturbative definition $S_i = \gs M_i$)
and $\delta \rho (\lam) = \delta S_i \, \delta(\lam-e_i)$.
As in sec.~\ref{U2metII}
(up to terms which will not affect our discussion)
\bea
{\partial \fre \over \partial S_i}
&=& \int_{I_2}^{I_1} u
= \int_{I_3}^{I_2} u
= \half \int_{I_3}^{I_1} u
\non\\
\frpp 
&=& -{\beta \over 2} \int_0^I u_1 
  = +{\beta \over 2} \int_0^I u_3 
  = -{\beta \over 4} \left[ \int_{0_1}^{I_1} u -  \int_{0_3}^{I_3} u \right]
\eea
where $I_i$ denotes infinity on the $i$th sheet, 
and $0_i$ is the point $z=0$ on sheet $i$. 

Next, we wish to extremize the effective superpotential  (\ref{Weffdef})
\be
\Weff 
= - \sumN  {\partial  \fre \over \partial S_i} - 4 \frpp
\ee
(setting $N_i=1$ and $N_0=0$).
By changing basis as in sec.~\ref{U2metII} and varying w.r.t~the coefficients 
of the arbitrary polynomials 
$r_1(z)$ and $t_1(z)$ in the matrix model curve (\ref{USMMcurvers}),
one obtains
\be
\label{SAAbel}
0 = (N-2 \beta) \int_{p_0}^{I_1} \zeta_k 
  - (N-2 \beta) \int_{p_0}^{I_3} \zeta_k 
  + 2\beta \int_{p_0}^{0_1} \zeta_k 
  - 2\beta \int_{p_0}^{0_3} \zeta_k
= 0 \quad {\rm (modulo\; period \;lattice)}\,.
\ee
This condition seemingly implies, 
by Abel's theorem, the existence of a function with a
pole of order $N-2\beta$ at $I_1$, 
a zero of order $N-2\beta$ at $I_3$,
a pole of order $2\beta$ at $0_1$, 
and a zero of order $2\beta$ at $0_3$, 
and regular everywhere else. However, there is one important caveat. In the 
undeformed ($r_1(z)=t_1(z)=0$) curve (\ref{USMMcurvers}), 
$z=0$ is a singular double-point. 
If a cut opens up between sheets one and three
when $r_1(z)$ and $t_1(z)$ are turned on,
then the points $0_1$ and $0_3$ will be identical 
and the last two terms in (\ref{SAAbel}) will not contribute, and the 
function will be regular at $z=0$ on all the sheets.

\medskip
\noindent {\bf (i)}  $\U(N) + \Ysymm$ ($\beta=1$)
\medskip

We will now show that this problem has a solution:
that for some choice of $r_1(z)$ and $t_1(z)$
in the matrix-model curve (\ref{USMMcurvers}),
there exists a function $f(z)$ on this curve 
with divisor implied by (\ref{SAAbel}) via Abel's theorem.  
For simplicity, we restrict ourselves to the special 
case in which $N$ is even and $\sigma_1=0$; 
these conditions are such as to make
$W'(z) - W'(-z)$ a polynomial of order $N-3$.
Consider $r_1(z)=0$ and $t_1(z) \propto z^2 (W'(z)-W'(-z))^2$,
i.e. the matrix-model curve\footnote{
The polynomial $t_1(z)$ is of degree $2N-2$ or less 
only when only when $N$ is even and $\sigma_1=0$.  
In the more general case, a more complicated choice
of $r_1(z)$ and $t_1(z)$ would be necessary.}
\be \label{repeat2}
(u-W'(z))\,u\,(u-W'(-z)) = {\rm~const} \times z^2 (W'(z)-W'(-z))^2\,.
\ee
The function
\be \label{USf}
f(z) = \frac{u-W'(-z)}{u-W'(z)}
\ee
defined on this curve,
has the following asymptotic behavior near $I_i$, infinity  
on the three sheets:
\be
I_1: \quad f(z) \sim  z^{N-2}\,, \qquad 
I_2: \quad f(z) \sim {\rm~const~} \,, \qquad
I_3: \quad f(z) \sim z^{-N+2} 
\ee
and the following behavior near $0_i$, $z=0$ on the three sheets:
\be
0_1: \quad f(z) \sim z^{-2} \,, \qquad 
0_2: \quad f(z) \sim {\rm~const~} \,, \qquad
0_3: \quad f(z) \sim z^{2}  
\ee
As in sec.~\ref{U2metII} 
one can show that $f(z)$ is regular everywhere else,
and so satisfies the conditions implied by (\ref{SAAbel}) and Abel's theorem.
Assuming that the solution is unique (which we have not been able to prove), 
we find that extremization of $\Weff$ (via Abel's theorem) 
implies a matrix-model curve of the form (\ref{repeat2}). 
Upon redefining $W'(z)  \propto P(z) -3 \LSp z^2 $,
this curve precisely agrees with the M-theory curve (\ref{USucurve}),
and the function (\ref{USf}) is proportional, up to a factor of $z^2$, 
to $y$ (\ref{USy}).

\medskip
\noindent {\bf (ii)}  $\U(N) + \Yasymm$ ($\beta=-1$)
\medskip

To show that the problem has a solution,
consider $r_1(z)=0$ and $t_1(z) \propto z^{-2} (W'(z)-W'(-z))^2$,
i.e. the matrix-model curve
\be \label{repeat3}
(u-W'(z))\,u\,(u-W'(-z)) = {\rm ~const} \times
\left(W'(z)-W'(-z)\over z\right)^2
\ee
For this curve, there is a cut opening up at $z=0$ 
between sheets 1 and 3, so that the last two terms 
in (\ref{SAAbel}) do not contribute. 
The function 
\be \label{UAf}
f(z) = \frac{u-W'(-z)}{u-W'(z)}
\ee
on the curve (\ref{repeat3})
has the following asymptotic behavior 
near $I_i$, infinity  on the three sheets:
\be
I_1: \quad f(z) \sim  z^{N+2}\,, \qquad 
I_2: \quad f(z) \sim {\rm~const~} \,, \qquad
I_3: \quad f(z) \sim z^{-N-2} 
\ee
and is regular near $z=0$ on all three sheets.  
As in sec.~\ref{U2metII} one can show that 
$f(z)$ is also regular everywhere else.
Thus, the function (\ref{UAf}) on the Riemann surface (\ref{repeat3}) 
has precisely the divisor specified by (\ref{SAAbel}).
Assuming that the solution is unique
(which we have not shown), we find that 
Abel's theorem implies a Seiberg-Witten curve 
of the form (\ref{repeat3}). 
Setting $W'(z) \propto P(z)$,
eq.~(\ref{repeat3}) precisely agrees with the M-theory curve (\ref{UAucurve}),
and the function (\ref{UAf}) is proportional, up to a factor of $z^2$, 
to $y$ (\ref{UAy}).

\setcounter{equation}{0}
\section{Seiberg-Witten differential from the matrix model}
\label{sSWdiff}

In this section
we will discuss the derivation of 
the Seiberg-Witten differentials
for the $\cN=2$ gauge theories described in section \ref{sMtheory} 
from the matrix model point of view.

As is the case for the SW curve, the SW differential can be determined
order-by-order in $\Lambda$ by using perturbative matrix
model calculations.   We will illustrate this in the first
part of this section, after which we will very briefly discuss some 
other approaches, such as the one pursued  
in (version 3 of) ref.~\cite{Naculich:2002b}. 

\subsection{Method I: perturbative determination of $\la_{SW}$}
\label{sLaSWpert}

On the first sheet of the Riemann surface, 
we have (for all the models)
\cite{Cachazo:2002a,Dijkgraaf:2002c, Dijkgraaf:2002d,Cachazo:2002b}
\be \label{laT}
\laSW = z \, T(z) \, \D z,
\qquad 
T(z) = \bvev{\tr\!\left( \frac{1}{z-\phi} \right) } 
= \sum_{n=0}^{\infty}z^{-n-1} \vev{\tr\,\phi^n}
\ee
where $\vev{\tr\,\phi^n}$
is the gauge-theory vev of the adjoint field 
and $T(z)$ is sometimes called $h(z)$.
The relation between gauge-theory and matrix-model vevs 
can be obtained using the methods 
in ref.~\cite{Naculich:2002a}
\renewcommand{\theequation}{\arabic{section}.\arabic{equation}\alph{abc}}
\setcounter{abc}{1}
\newcounter{bc}
\setcounter{bc}{3}
\be \label{U2vevrel}
\vev{\tr \, \phi^n} = 
\sumN \left[\frac{\pa}{\pa S_i} + \frac{\pa}{\pa \hS_i}\right] 
\, g_s \, \vev{\tr \, \Phi^n}_{S^2}
\ee
\addtocounter{equation}{-1}\addtocounter{abc}{1} 
\renewcommand{\theequation}
{\arabic{section}.\arabic{equation}\alph{abc},\alph{bc}}
\be
 \label{USvevrel}
\vev{\tr \, \phi^n} = 
\sumN \frac{\pa}{\pa S_i}  
\, g_s \, \vev{\tr \, \Phi^n}_{S^2} + 4 \vev{\tr\,\Phi^n}_{\sR\sPP^2} 
\ee
\renewcommand{\theequation}{\arabic{section}.\arabic{equation}}where 
the first equation is for model (a), 
$\U(N){\times}\U(N)$ with a bifundamental hypermultiplet,
and the second equation is for models (b) or (c),
$\U(N)$ with a symmetric or antisymmetric hypermultiplet, respectively.
It is understood that 
the rhs is evaluated at $S_i = \vev{S_i}$ 
(as well as $\hS_i = \vev{\hS_i}$ for model (a) 
and $S_0 = \vev{S_0}$ for model (c)).
Since we are only interested in the first two orders 
in perturbation theory, we may write \cite{Naculich:2002a}
\be
\vev{\tr\Phi^n} = \sumN 
\left[M_i e_i^n + n \,e_i^{n-1} \vev{\tr\,\Psi_{ii}} 
+ \half n(n{-}1) e_i^{n-2} \vev{\tr\, \Psi_{ii}^2} + \ldots\right] \,.
\ee
For model (c)  we also have the extra 
terms\footnote{Note that, for this model,
$\vev{\tr\Psi_{00}}\equiv 0$ because of the Sp-condition on $\Phi_{00}$.} 
\be
M_0  \de_{n,0} 
+ \vev{\tr\, \Psi_{00}^2} \de_{n,2} + \ldots 
\ee
It follows from these expressions that for all three models, 
the leading term in the perturbative expansion for $T(z)$ is given by 
\be
T(z)_{\rm pert} = \sumN \frac{1}{z-e_i} \,.
\ee

The matrix model expectation values $\tadpole$ and $\rptadpole$
are given in eqs.~(\ref{tadpoleUU}), (\ref{tadpoleSA}), (\ref{rptadpoleSA})
of sec.~\ref{sPert},
and for all three models,  we have
\be
\vev{\tr\, \Psi_{ii}^2}_{S^2} = {\gs\over \al} \sumN {M_i^2 \over R_i},
\qquad \vev{\tr\, \Psi_{ii}^2}_{\sR \sPP^2}=0 \,.
\ee
In addition, for model (c), we will need\footnote{Note
that $\vev{ \tr \Psi_{00}^2 }_{S^2} 
\propto S_0^2$ does not contribute to (\ref{USvevrel}) because 
of the absence of a derivative w.r.t.~$S_0$. }
\be
\vev{\tr\, \Psi_{00}^2}_{\sR \sPP^2}=
\frac{1}{2\al} \frac{S_0}{R_0 } .
\ee
Using these results together with the above formul\ae{} leads 
to the following expressions for the one-instanton 
corrections (see section \ref{sPert} for details about the notation)
\renewcommand{\theequation}{\arabic{section}.\arabic{equation}\alph{abc}}
\setcounter{abc}{1}
\bea \label{U2TLa}
T(z)_{\rm 1-inst}
&=& 
\La^N 
\sumN \bigg\{ 
{1 \over (z-e_i)^2} 
\bigg[ {2 \over R_i} \sum_{j\neq i}\frac{\Tj }{R_j \eij} 
       + {1 \over R_i} \sum_{j} \frac{\hTj }{\hR_j \hij} 
       - {\Ti  \over R_i^2}\sum_{j}\frac{1}{\hij} \bigg] \non \\
&& \qquad\;\; + \, {2 \over (z-e_i)^3} 
\frac{T_i}{R_i^2} \bigg\}
\\[3pt]\addtocounter{equation}{-1}\addtocounter{abc}{1}  \label{USTLa}
T(z)_{\rm 1-inst} 
&=& 
\La^{N-2} 
\sumN \bigg\{ {1 \over (z-e_i)^2} 
\bigg[ {2 \over R_i} \sum_{j\neq i}\frac{\Gj }{R_j \eij} 
       - {1 \over R_i} \sum_{j} \frac{\Gj }{R_j \gij} 
       - {\Gi  \over R_i^2}\sum_{j}\frac{1}{\gij} 
	- {2 \Gi \over R_i^2 e_i} \bigg] \non \\
&& \qquad\quad\;\; +\, {2 \over (z-e_i)^3} 
\frac{G_i}{R_i^2} \bigg\}
\\[3pt] \addtocounter{equation}{-1}\addtocounter{abc}{1} \label{UATLa}
T(z)_{\rm 1-inst} 
&=& 
\La^{N+2} \sumN \bigg\{ {1 \over (z-e_i)^2} 
\bigg[ {2 \over R_i} \sum_{j\neq i}\frac{\Gj }{R_j \eij} 
       - {1 \over R_i} \sum_{j} \frac{\Gj }{R_j \gij} 
       - {\Gi  \over R_i^2}\sum_{j}\frac{1}{\gij} 
+ {2 \Gi \over R_i^2 e_i}  \non \\
&& \qquad \qquad +\, {2 (-1)^N \over R_i e_i} \sum_k \frac{1}{e_k}\bigg] 
+ \, {2 \over (z-e_i)^3} 
\frac{G_i}{R_i^2} 
- \frac{4}{z^3} \frac{1}{\prod_j e_j}\bigg\}
\eea
\renewcommand{\theequation}{\arabic{section}.\arabic{equation}}where 
eqs. (\ref{U2TLa}), (\ref{USTLa}), and (\ref{UATLa})
correspond to models (a), (b), and (c), respectively.
Next we use the identities 
(\ref{UUident}), (\ref{symident}), and (\ref{antiident}), 
together with the definitions
(\ref{eeprimeUU}), (\ref{eprimesym}) to obtain
\be
T(z)\D z = \D \log P(z) - \La^{N-2\beta} \D K(z)  + \cO( \La^{2N-4\beta})
\ee
with  
\renewcommand{\theequation}{\arabic{section}.\arabic{equation}\alph{abc}}
\setcounter{abc}{1}
\bea \label{U2K}
\beta=0,\qquad &&K(z) = \frac{\hP(z)}{P(z)^2}
\\[3pt]\addtocounter{equation}{-1}\addtocounter{abc}{1}  \label{USK}
\beta=1,\qquad &&K(z) = (-1)^N \frac{z^2 P(-z)}{P(z)^2}
\\[3pt] \addtocounter{equation}{-1}\addtocounter{abc}{1} \label{UAK}
\beta=-1,\qquad && K(z) = (-1)^N \left[
\frac{1}{z^2}\frac{P(-z)}{P(z)^2} 
- \frac{3}{z^2}\frac{1}{P(z)} \right]
\eea
\renewcommand{\theequation}{\arabic{section}.\arabic{equation}}recalling 
that
$P(\x) = \prod_{i=1}^N (\x-e'_i)$ and $\hP(\x) = \prod_{i=1}^N (\x-\he'_i)$.

These results are consistent with, using (\ref{laT}), what one  
obtains by expanding the M-theory result (\ref{laSWm}) 
on the first sheet of the Riemann surface given by
(\ref{U2y})-(\ref{UAy}).

\subsection{Other methods}

For the $\UxU$ gauge theory with a bifundamental hypermultiplet, 
we may combine eqs.~(\ref{laT}), (\ref{U2vevrel}), and (\ref{ress})
to derive the relation (on sheet one of the Riemann surface)\footnote{
In the equations in this section, it is understood that, 
after taking the derivatives, 
the results are to be 
evaluated at $S_i =\vev{S_i}$ and $\hS_i = \vev{\hS_i}$.}
\be  
\label{U2laSW}
\laSW = z \, T(z) \, \D z, \qquad
T(z) = \sumN \left[ \frac{\pa}{\pa S_i} + \frac{\pa}{\pa \hS_i}\right] 
\, \om(z) 
\ee
where $\om(z)$ is the leading (sphere) part of the resolvent.
Similarly, for the $\U(N)$ gauge theory with 
one symmetric or antisymmetric hypermultiplet,
we may combine eqs.~(\ref{laT}), (\ref{USvevrel}), and (\ref{res})
to derive (on sheet one)
\be  \label{SAT}
\laSW = z \, T(z) \, \D z, \qquad
T(z) = 
\sumN \frac{\pa}{\pa S_i} \, \om_0(z)  + \om_{1/2}(z)
\ee
where $\om_0(z)$ is the leading (sphere) part of the resolvent
and $\om_{1/2}(z)$ is the subleading ($\R\PP^2$) part.
These results appeared (using a different approach) 
in \cite{Naculich:2003,Kraus:2003b}),
and earlier for the case of $\U(N)$ without \cite{Gopakumar:2002}
and with \cite{Naculich:2002b} fundamental hypermultiplets.
Now we may use eq.~(\ref{U2uis}) to show, for $\UxU$,
\be
\label{noname}
\laSW = -z\, \sumN \left[\frac{\pa }{\pa S_i} 
+ \frac{\pa }{\pa \hS_i}\right] u \, \D z \,.
\ee
Since both $\la_{SW}$ and $u$ are defined on all the sheets,
this equation extends to the entire Riemann surface. 
A similar equation may be derived for $\U(N)$ with a symmetric
or antisymmetric  hypermultiplet starting from (\ref{SAT}).

Two methods for computing $\laSW$ 
(in addition to the approach used in sec.~\ref{sLaSWpert})
present themselves. 
First, one may use perturbation theory to calculate 
the curve polynomials $r_1(z)$ and $t_1(z)$ 
as functions of the $S$'s, 
and then use the curve equation to calculate the derivatives of $u$
in eq.~(\ref{noname}). 
In appendix \ref{aUNf},
we use the analog of this approach
to determine the one-instanton contribution to
$T(z)$ for the $\U(N)$ model with $N_f<N$ fundamentals.
Obtaining the one-instanton contribution for the theories considered
in this paper is straightforward, but somewhat cumbersome, 
so we will not pursue it here.

A second approach is to investigate the integrals 
around the A-cycles together with 
the behavior at infinity, and try to use this 
information to pin down the function $T(z)$.
This method was used in (version 3 of) \cite{Naculich:2002b} 
to determine $T(z)$ for the $\U(N)$ model with $N_f<N$ fundamentals.
For that case $T(z)$ was given by $\frac{\D \psi}{\psi}$,  
where $\psi(z)$ was the function arising from Abel's theorem.
Such a relation, involving the function $f(z)$ given in (\ref{U2f}),
probably also holds for the $\UxU$ theory.
Recalling that $f(z)$  (\ref{U2f}) is proportional to $y$ (\ref{U2y}),
using (\ref{UUPW}), this implies $T(z) = \frac{\D y}{y}$, 
which is consistent with the M-theory result (\ref{laSWm}). 
For the $\U(N)$ models the situation is less clear since $f(z)$ 
(\ref{USf}), (\ref{UAf}) and $y$ (\ref{USy}), (\ref{UAy}) differ 
by rescalings with $z^2$.

\section{Summary}

This paper is part of a larger program aimed at studying the 
applicability of the Dijkgraaf-Vafa matrix model approach 
to $\cN=2$ supersymmetric gauge theory theories, i.e.~those theories 
amenable to the methods of Seiberg-Witten theory. 
Previously it has been shown that one can recover the ingredients  
of the Seiberg-Witten solution from the matrix model for theories 
with hyperelliptic Seiberg-Witten curves.
In this paper 
we focused on three different models: (a) $\U(N){\times}\U(N)$ 
with matter in the bifundamental representation,
(b) $\U(N)$ with matter in the symmetric representation, and 
(c) $\U(N)$ with matter in the antisymmetric representation. 
Each of these theories is described by a 
cubic non-hyperelliptic Seiberg-Witten curve.
Our goal was to determine the Seiberg-Witten curve and differential, 
as well as the order parameters and the prepotential, for the above models, 
entirely within 
the context of the matrix model, without reference to string/M-theory.
Our results confirm the results previously obtained using M-theory.

For models (a) and (b), a straightforward generalization of our earlier work
(including a refinement of the prescription for $\tau_{ij}$, as
discussed in \cite{Naculich:2003}) produced expressions in complete 
agreement with earlier results in the literature. 
For model (c), $\U(N)$ with matter in the antisymmetric representation, 
the naive extension of earlier work leads to 
discrepancies with previous results in the literature. 
The discrepancies 
occur when one expands around the simplest matrix model vacuum with 
$\prod_i \U(M_i)$ gauge symmetry, where each $M_i\rar \infty$ with 
$S_i=\gs M_i$ fixed. 
However there are other vacua; one of these has 
$\Sp(M_0){\times}\prod_i \U(M_i)$ gauge symmetry \cite{Klemm:2003}, 
and hence an additional parameter $S_0 =\gs M_0$. 
When  one expands around this vacuum, extremization 
of $\Weff$ (with $N_i=1$ and $N_0=0$)
leads to non-zero vevs for both $S_i$ and $S_0$, and all 
discrepancies are removed.
What is missing is a guiding principle for the enlargement of the 
matrix-model vacua for the $\Yasymm$ theory.
(A similar enlargement of the vacua of the $\Ysymm$ theory 
seemingly leads to an inconsistency.)

We close with an empirical observation
about the form of the one-instanton contribution
to the prepotential.
For pure $\U(N)$ gauge theory \cite{Naculich:2002a},
or $\U(N)$ with matter in the $\Yfund$ \cite{Naculich:2002b}
or $\Ysymm$ (sec.~\ref{sSAperttau}) representations, 
one may verify, using our results, 
that the following equation holds
\be
\label{universal}
2 \pi i \cF_{1-{\rm inst}} = \sum_i { \vev{S_i} \over W''(e_i) }
\ee
where the sum is over all the extrema of the superpotential $W(z)$.
Observe that the expression (\ref{universal}) has a finite limit
when the coefficient $\al$ multiplying the superpotential 
is taken to zero to restore $\cN=2$ supersymmetry.
For $\UxU$ with a bifundamental hypermultiplet (sec.~\ref{sUUperttau}), 
the sum also extends over the extrema of $\hW(z)$:
\be
2 \pi i \cF_{1-{\rm inst}} = \sum_i { \vev{S_i} \over W''(e_i) }
+ \sum_i { \vev{\hS_i} \over -\hW''(\he_i) }
\ee
where the relative minus sign is due to the fact that
$\hW(z)$ enters with a minus sign in the superpotential (\ref{U2suppot}).
Finally, for $\U(N)$ with $\Yasymm$ (sec.~\ref{sSAperttau}), 
there is an additional contribution  from the vacuum state at $z=0$:
\be
2 \pi i \cF_{1-{\rm inst}} = \sum_i { \vev{S_i} \over W''(e_i) }
                                  + { \vev{S_0} \over W''(0) }
\ee
However, at present we do not have an understanding of why these 
results are true.

\section*{Acknowledgments}

We are grateful to F. Cachazo for helpful discussions.
HJS would like to thank the string theory group and Physics 
Department of Harvard University for their hospitality extended 
over a long period of time.

\appendix
\renewcommand{\theequation}{\Alph{section}.\arabic{equation}}

\section*{Appendices}
\setcounter{equation}{0}
\section{{\large $\U(N) +N_f\,$}$\protect\Yfund$ curve 
and {\large $T(z)$} from perturbation theory} \label{aUNf}
In ref.~\cite{Naculich:2002b}, we derived the SW curve for
the $\cN=2$ $\U(N)$ gauge theory with $N_f$ fundamental multiplets
from the associated matrix model.
Specifically, the saddle-point solution of the matrix model
gives rise to the equation for the resolvent\footnote{ 
We have converted to the notation in this paper; 
to convert back let $\om(z) \to - S \om(z)$. }
\be
\label{resolventeq}
\om^2(z) - W' (z) \, \om(z) + {\ts \frac{1}{4}} f(z) = 0
\ee
where $W'(z) = \al \prodN (z-e_i)$ and using (\ref{cndef})
\be
f(z)
= 4 \gs \bvev{\tr\!\left(\frac{W'(z)-W'(\Phi)}{z-\Phi}\right) } 
= 4 \sum_{n=0}^{N-1} c_n(z) \, \gs \vev{\tr(\Phi^n)} 
\ee
is an $(N-1)$th order polynomial. 
Defining 
$y(z)  = - 2 \om(z) + W' (z)$
one may rewrite eq.~(\ref{resolventeq}) as a hyperelliptic curve
\be
\label{hyper}
y^2 =  W' (z)^2 - f(z)\,. 
\ee 

In ref.~\cite{Naculich:2002b}, we determined the
polynomial $f(z)$ using Abel's theorem
(see also \cite{Cachazo:2002a}).
In this appendix, we will show how this polynomial can be evaluated
using the perturbative solution.
The lowest order contribution to the
matrix model vev is \cite{Naculich:2002a,Naculich:2002b}
\be \label{urgug}
\gs \vev{\tr(\Phi^n)}= 
\al \La^{2N-N_f} \sum_i {L_i \over R_i}  e_i^n  + \cO(\La^{4N-2N_f})
\ee
where $L_i = \prod_{I=1}^{N_f}(e_i+m_I)$ and $R_i = \prod_{j\neq i} (e_i-e_j)$. 
It follows from (\ref{urgug}) that 
\be
f(z)=4\al\,\La^{2N-N_f} \left[W'(z)-W'(e_i) \right] \sum_i {L_i \over R_i(z-e_i)} 
+ \cO(\La^{4N-2N_f})\,.
\ee
Now using the fact that $W'(e_i)=0$ together with 
\be
\sum_i {L_i \over R_i (z-e_i)}
=  {\prod_{I=1}^{N_f} (z+m_I)\over \prodN (z-e_i)} - \poly 
\ee
where $\poly$ is the polynomial part of 
$\prod_{I=1}^{N_f} (z+m_I)/ \prodN (z-e_i) $,
we finally obtain
\be
\label{exactresult}
f(z) = 4 \, \al^2 \La^{2N-N_f} 
\left( \prod_{I=1}^{N_f} (z+m_I) - \poly \prodN (z-e_i)   \right)
+ \cO(\La^{4N-2N_f})
\ee
which is precisely  the result obtained by Abel's theorem in
ref.~\cite{Naculich:2002b}.

When $N_f<N$, the polynomial $\poly$ vanishes. 
Moreover, in that case, eq.~(\ref{exactresult}) 
is exact to all orders in $\La$ \cite{Naculich:2002b}, 
which points to the existence of a non-renormalization theorem. 

Similarly, in (version 3 of) \cite{Naculich:2002b} 
(see also \cite{Gopakumar:2002}), 
we derived the expression for the Seiberg-Witten differential
for this theory (with $N_f < N$)
from the saddle-point approach
\be
\laSW = z \, T(z) \, \D z, \qquad
T(z) = - \half \sumN {\partial y \over \partial S_i} \vevS
         + {y - W'(z) \over 2 y } {f'(z) \over f(z)} \,.
\ee
Let us now evaluate this expression using only the
perturbative solution.   
As in eq.~(\ref{U2result}), we have
\be
f(z) = 4 W'(z) \sum_i {S_i \over z-e_i}  + \cO(S_i^2)
\ee
so that, using eq.~(\ref{hyper})
\be
-\half \sumN {\partial y \over \partial S_i}  
= {1 \over 4y} \sumN {\partial f \over \partial S_i}
= \sumN {W'(z)\over y (z-e_i) }  + \cO(S_i)
= {W''(z)\over y }  + \cO(S_i)\,.
\ee
Evaluating this expression at $S_i = \vev{S_i}$,
the $\cO(S_i)$ term is subleading in $\La$,
so the leading order contribution to $T(z)$ is
\be
T(z) = {1\over y} \left[ W''(z) + \half (y - W'(z)) {f' \over f} \right]\,.
\ee
However, as we showed in \cite{Naculich:2002b},
this expression is exact, 
again pointing to the existence of a non-renormalization theorem 
(when $N_f < N$).

\setcounter{equation}{0}
\section{Some technical details} \label{appdet}
In this appendix we collect some technical details of the perturbative calculations performed in section \ref{sPert}.

\subsection{{\large $\U(N){\times}\U(N)$} }

The integration over the quadratic fields 
of the matrix-model partition function (\ref{ZUU}) yields
(up to an $e_i$-independent quadratic monomial in the $S_i$, $\hS_i$'s) 
\bea
\label{freeenergytwoUU}
\fre (S,\hS)
& =& 
- \sumN S_i W(e_i) 
+ \half \sumN S_i^2 \log \left( S_i\over \alpha R_i \La^2 \right) 
+ \sumN \sum_{j\neq i}  S_i S_j \log\left(\eij \over \La \right) +  \non\\
&& 
+ \sumN \hS_i \hW(\he_i) 
+ \half \sumN \hS_i^2 \log \left( \hS_i\over -\alpha \hR_i \La^2 \right) 
+ \sumN \sum_{j\neq i}  \hS_i \hS_j \log\left(\heij \over \La \right) +  \non\\
&&
- \sum_i  \sum_j  S_i \hS_j \log\left(\hij \over \La \right) + 
\sum_{n \ge 3} \fren (S,\hS)
\eea
where $\eij = e_i-e_j$, $\hij = e_i-\he_j$  
and $R_i = \prod_{j \neq i} \eij$.  
The term $\fren (S,\hS)$ is an $n$th order polynomial in $S_i$ and $\hS_i$
arising from planar loop diagrams built from the interaction
vertices.
The contribution to $\fre (S,\hS)$ cubic in $S_i$ and $\hS_i$,
\bea
\label{freeenergythreeUU}
\alpha \frethree  (S, \hS)
&=& 
\Bigg[ {\ts \frac{2}{3}}
\sum_i  \frac{S_i^3}{R_i} \left( \sumk {1 \over \eik} \right)^2
- {\ts \frac{1}{4} }
\sum_i \frac{S_i^3}{R_i} \sumk \suml {1 \over \eik \eil} 
- 2 \sum_i \sumk \frac{S_i^2 S_k}{R_i \eik} \sum_{\ell \neq i} \frac{1}{\eil}  
\non\\ &&
+\, 2 \sum_i \sumk \sum_{\ell \neq i} \frac{S_iS_kS_\ell}{R_i \eik \eil}
- \sum_i \sumk  \frac{S_i^2 S_k}{R_i \eik^2}  
+ \sum_i \sum_k \frac{S_i^2 \hS_k}{R_i \hik} \sum_{\ell \neq i} 
\frac{1}{\eil}  + \half \sum_i \sum_k  \frac{S_i^2 \hS_k}{R_i \hik^2} 
 \non\\ &&
- \, 2 \sum_i \sumk \sum_{\ell} \frac{S_iS_k\hS_\ell}{R_i \eik \hil} 
+ \half \sum_i \sum_k \sum_{\ell} \frac{S_i\hS_k\hS_\ell}{R_i \hik \hil}
\Bigg]
 - \Bigg[ S \leftrightarrow \hS; 
e_i \leftrightarrow \he_i \Bigg]\,,
\eea
is obtained by adding to the result from ref.~\cite{Naculich:2002a} 
the planar two-loop diagrams containing  $B$ loops that can
be drawn on a sphere.

Using the above expressions to calculate (\ref{WeffdefUU}) 
and extremizing (\ref{WextremeUU}) leads to
\bea
\label{SvevUU2}
\vev{S_i} &=& 
{\alpha \Ti \over R_i}  \LU 
\Bigg[ 1 + \LU \Bigg\{ 
\sum_{\ell}  \sumk \left(
- {2 \Ti \over R_i^2 \eik \hil} 
+ {2 \Tk \over R_i R_k \eik \hil} 
- {2 \Tk \over R^2_k \eik \hkl} 
+ {2 \hTl \over R_k \hR_\ell \eik \hkl} 
\right)
\non \\ &&
\!\!\!\!\! +\,
\sumk \suml  \left(
{3 \Ti \over 2 R_i^2 \eik\eil} + {4 \Tl  \over R_k R_\ell \eik \ekl} 
\right)
+ \sumk \left(
\frac{2 \Ti}{R_i^2\eik^2} 
- \frac{4\Ti}{R_i R_k \eik^2}
+ \frac{2 \Tk }{R_i R_k \eik^2}
+ \frac{2 \Tk}{R_k^2\eik^2}
\right) 
\non \\ &&
\!\!\!\!\! +\, \sum_{\ell}  \sum_{k \neq \ell} 
\left(  {2 \hTk \over \hR_k \hR_\ell \hekl \hil}  \right) 
+ \sum_k \sum_{\ell}  \left(
  { \hTk \over R_i \hR_k \hik \hil} 
- { \hTk \over \hR^2_k  \hik \hlk} 
+ { \Tk \over R_k \hR_\ell \hil \hkl} 
\right)
\non\\ &&
\!\!\!\!\! + \, \sum_k \left(
- { \Ti \over R_i^2 \hik^2} 
- { \hTk \over \hR_k^2 \hik^2} 
+ { \hTk \over R_i \hR_k \hik^2} 
\right)
\Bigg\} + \cO(\La^{2N}) \Bigg] 
\non\\
\vev{\hS_i} &=& 
 -  {\alpha \hTi \over \hR_i}  \LU \Bigg[ 1 + 
\LU \Bigg\{  {\rm above~expression~with~} e_i \leftrightarrow \he_i \Bigg\} 
+ \cO(\La^{2N}) \Bigg] 
\eea
where $\Ti = \prod_{j=1}^N (e_i - \he_j)$,
$\hTi = \prod_{j=1}^N (\he_i - e_j)$,
$\hR_i = \prod_{j \neq i} (\he_i-\he_j)$,
and  various constants 
have been absorbed into a redefinition of the cut-off $\La$.

The gauge-coupling matrix (\ref{UUtauij}) can be expanded as
\be \label{UUtauexp}
\tau_{ij}   =
\tau_{ij}^{\rm pert} + \sum_{d=1}^\infty \La^{Nd} 
\tau_{ij}^{(d)}\,
\ee
and similarly for $\tau_{\hi\hj}$ and $\tau_{i \hj}$.
Using (\ref{freeenergytwoUU}), (\ref{freeenergythreeUU}), (\ref{SvevUU2}),
and then reexpressing the result in terms of $a_i$ and $\ha_i$  
using (\ref{aeUU}),
the first two terms in the expansion (\ref{UUtauexp}) can be determined.
The perturbative contribution is (up to additive constants) 
\be \label{UUtaupert}
2\pi i \tau^{\rm pert}_{ij} 
=
\delta_{ij} 
\Bigg[ - 2 \sumk \log \left( a_i - a_k  \over \La \right)
       +   \sum_k \log \left( a_i - \ha_k  \over \La \right)
\Bigg]  
 + (1 - \delta_{ij}) 
\Bigg[ 2  \log \left( a_i - a_j  \over \La \right) \Bigg] 
\ee
and the one-instanton contribution, after some algebra,  is
\bea \label{UUtauone}
2 \pi i \tau^{(1)}_{ij}  \!\!&=&\!\!
\de_{ij} \Bigg[ 
\sumk \left(
  {6\Tk \over R_k^2 \aik^2}  
+ {2\Ti \over R_i^2 \aik^2}
+ {4 \Ti \over R_i^2 \aik} \sum_{\ell \neq i}  {1 \over \ail}
\right) 
- \sum_\ell {\Ti \over R_i^2 \hil} 
\left(  \sumk  {4  \over  \aik} - \sum_k  {1 \over  \hik} +  {1 \over \hil} 
\right) \! \Bigg]  
\non \\ &&
\!\!\!\!\!\!\!\!\!\!\!\!\!\!\!\!\!\!\!\!\!\!\!\!\!\!\!\!\! 
+ \, (1{-}\de_{ij}) \Bigg[ 
\sum_{k\neq i,j} \!
{ 4 \Tk \over R_k^2  \aik\ajk} 
+ \sum_{k} \!
{ \hTk \over \hR_k^2 \hik\hjk} 
- \bigg[ {2\Ti \over R_i^2 \aij} \!
\left(\sumk {2 \over \aik} - \sum_k {1 \over \hik} + {1 \over \aij} \right)
+ ( i \leftrightarrow j) \bigg] \!
\Bigg]  
\eea
where now 
$\aij= a_i - a_j$, 
$\hij= a_i - \ha_j$, 
$R_i = \prod_{j \neq i} \aij$,
$\hR_i = \prod_{j \neq i} (\ha_i-\ha_j)$,
$\Ti = \prod_{j=1}^N (a_i - \ha_j)$, and
$\hTi = \prod_{j=1}^N (\ha_i - a_j)$.
The expressions for $\tau_{\hi \hj}$ are obtained from those above
by letting $a_i \leftrightarrow \ha_i$.
Finally,
\bea \label{taumixed}
 2 \pi i \tau^{\rm pert}_{i \hj} &=&
-   \log \left( a_i - \ha_j  \over \La \right)  
\\
 2 \pi i \tau^{(1)}_{i \hj} &=&    
{\Ti \over R_i^2 \hij} 
\left(  \sumk {2 \over \aik}  - \sum_k {1 \over \hik} + {1 \over \hij}\right)
+ 2 \sumk \sum_{\ell} {\Tk \over R_k^2 \aik \hkl} 
+ \bigg(i \leftrightarrow j; a_i \leftrightarrow \ha_i \bigg) 
\non
\eea
These expressions may be expressed succinctly as the second derivative
of the prepotential (\ref{UUpre}).

\subsection{ {\large $\U(N)$} with $\protect\Ysymm$ or  $\protect\Yasymm$ }

Using the Feynman rules, the cubic contribution to the sphere 
part of the free energy can be shown to be
\bea
\label{freeenergythree}
\alpha \freethree &=& 
{\ts \frac{2}{3}}
\sum_i  \frac{S_i^3}{R_i} \left( \sumk {1 \over \eik} \right)^2
- {\ts \frac{1}{4} }
\sum_i \frac{S_i^3}{R_i} \sumk \suml {1 \over \eik \eil} 
- 2 \sum_i \sumk \frac{S_i^2 S_k}{R_i \eik} \sum_{\ell \neq i} \frac{1}{\eil}  
\non\\ &&
+ \,2 \sum_i \sumk \sum_{\ell \neq i} \frac{S_iS_kS_\ell}{R_i \eik \eil}
- \sum_i \sumk  \frac{S_i^2 S_k}{R_i \eik^2} 
+ \sum_i \sum_k \frac{S_i^2 \mS_k}{R_i \gik} \sum_{\ell \neq i} \frac{1}{\eil}  
\non\\ &&
- \,2 \sum_i \sumk \sum_{\ell} \frac{S_iS_k\mS_\ell}{R_i \eik \gil}
+ \half \sum_i \sum_k \sum_{\ell} \frac{S_i\mS_k\mS_\ell}{R_i \gik \gil}
+ \half \sum_i \sum_k  \frac{S_i^2 \mS_k}{R_i \gik^2} \non \\
&& 
+\, 
2 \sum_i \sum_{j\neq i} \frac{ S_0 S_i S_j}{R_i e_i \eij}
- \sum_{i,j} \frac{S_0 S_i \mS_j}{R_i e_i \gij} 
-\,\sum_i \sum_{j\neq i}\frac{S_0 S_i^2}{R_i e_i \eij} 
- \half \sum_i \frac{S_0 S_i^2}{R_ie_i^2}
\non \\ 
&& + \fourth \sum_{i,j}\frac{S_0 \mS_i(\mS_j-S_j)}{R_0 e_ie_j}
\;  + S_0^2 S_i {\rm ~terms~} + S_0^3 {\rm~ terms~}
\eea
where $\eij = e_i-e_j$, $\gij = e_i+e_j$,
$R_i = \prod_{j \neq i} \eij$, and 
$R_0 = - \prod_i(-e_i) \sum_j (1/e_j)$.  
This result was obtained by adding to 
the result from ref.~\cite{Naculich:2002a} 
the new planar two-loop diagrams that can
be drawn on a sphere and contains  $Y$ and ghost loops.
Similarly the quadratic contribution to the $\R\PP^2$ part of 
the free energy can be shown to be
\bea 
\label{frpquad}
\al \frptwo &=& \beta
\sum_i \left[
\half {S^2_i \over R_i e_i}  \sum_{\ell\neq i} \frac{1}{\eil}
- \sum_{j\neq i} \frac{S_iS_j}{R_i e_i \eij } 
+ \half \sum_j \frac{S_iS_j}{R_i e_i \gij} 
+ \fourth \frac{S^2_i}{R_i e_i^2}
- \half {S_i S_0 \over R_i e_i^2} 
+ \fourth {S_i S_0 \over R_0  e_i^2}   \right]
\non \\ && 
+ S_0^2 {\rm ~terms} \,,
\eea
The terms in (\ref{freeenergythree}) and (\ref{frpquad})
involving $R_0$ come from diagrams containing $\Psi_{00}$ legs.  
Since $\Psi_{00}\in\spl(M_0)$ for $\beta=-1$,    one must use 
$ (1/2 R_0) \left( \delta_d^a \delta_b^c + J^{ac} J_{bd} \right) $
for the propagator \cite{Cvitanovic:1976}.
Observe that eqs.~(\ref{freeenergythree}) and (\ref{frpquad})
obey the relation given in footnote \ref{drfoot}. 

Using the above expressions to calculate (\ref{Weffdef}) and 
then (\ref{Wextreme}), the solution for $\vev{S_i}$ 
can be evaluated in an expansion in $\La$ 
\bea
\label{Svev2}
\vev{S_i} &=& 
{\alpha \Gi \over R_i}  \LX 
\Bigg[ 1 + \LX \Bigg\{
\sumk \suml  \left(
{3 \Gi \over 2 R_i^2 \eik\eil} + {4 \Gl  \over R_k R_\ell \eik \ekl} 
\right)- \sum_{\ell}  \sum_{k \neq \ell} 
\left( {2 \Gk \over R_k R_\ell \ekl \gil}  
\right)
\non\\
&& 
+ \sumk \left(
\frac{2 \Gi}{R_i^2\eik^2} 
- \frac{4\Gi}{R_i R_k \eik^2}
+ \frac{2 \Gk }{R_i R_k \eik^2}
+ \frac{2 \Gk}{R_k^2\eik^2}
\right)  
- \sum_k \left(
 { \Gi \over R_i^2 \gik^2} 
+ { \Gk \over R_k^2 \gik^2} 
+ { \Gk \over R_i R_k \gik^2} 
\right)
\non\\
&&
+ \sum_{\ell}  \sumk \left(
- {2 \Gi \over R_i^2 \eik \gil} 
+ {2 \Gk \over R_i R_k \eik \gil} 
- {2 \Gk \over R^2_k \eik \gkl} 
- {2 \Gl \over R_k R_\ell \eik \gkl} 
\right) - \beta {2 \Gi \over R_i^2 e_i^2 } 
\non\\
&&
- \sum_k \sum_{\ell}  \left(
 { \Gk \over R_i R_k \gik \gil} 
+ { \Gk \over R^2_k  \gik \glk} 
+ { \Gk \over R_k R_\ell \gil \gkl} 
\right)
- \beta \sum_k \left(
  {2 \Gk \over R_i R_k \gik e_i} 
+ {2 \Gk \over R^2_k \gik e_k}  \right) 
\non\\
&&
+ 2 \delta_{\beta,-1}
(-1)^N \sum_j {\frac{1}{e_j}} 
  \left(
{1 \over R_0 e_i^2} - {1\over R_i e_i^2}  
+ \sumk {2 \over R_k e_k \eik}   
+ \sum_k {1\over R_k e_k \gik} 
+ \sum_k {1 \over R_i e_i \gik}
 \right)
\non\\ &&
+ \beta \sumk \left(
- {4 \Gi \over R_i^2 \eik e_i} 
+ {4 \Gk \over R_i R_k \eik e_i} 
- {4 \Gk \over R^2_k \eik e_k}  \right)
    \Bigg\} 
\eea
where 
$\Gi = e_i^{2\beta} \prod_{j} (e_i + e_j)$.
To evaluate (\ref{tauAS}) perturbatively as
\be
\label{tau_of_a}
\tau_{ij}   =
\tau_{ij}^{\rm pert} + \sum_{d=1}^\infty \La^{(N-2\beta)d} 
\tau_{ij}^{(d)}\,,
\ee
we use eqs.~(\ref{freeenergytwo}) and (\ref{freeenergythree})
in eq.~(\ref{tauAS}),
evaluate the resulting expression using eqs.~(\ref{S0vev}) 
and (\ref{Svev2}),
and then use the results of section \ref{ASae} 
to re-express the entire expression in terms of $a_i$. 
The perturbative contribution is (up to additive constants) 
\bea \label{SAtaupert}
2\pi i \tau^{\rm pert}_{ij} 
&=&
\delta_{ij} 
\Bigg[ - 2 \sumk \log \left( a_i - a_k  \over \La \right)
       +   \sumk \log \left( a_i + a_k  \over \La \right)
       + 2 (1+\beta) \log \left( a_i \over \La \right) 
\Bigg]  \non\\
&& + (1 - \delta_{ij}) 
\Bigg[ 2  \log \left( a_i - a_j  \over \La \right)
       +  \log \left( a_i + a_j  \over \La \right)
\Bigg] 
\eea
and the one-instanton contribution, after some algebra,  is
\bea \label{SAtauone}
2 \pi i \tau^{(1)}_{ij} &=&
\de_{ij} \Bigg[ \frac{2(\beta-1)}{a^2_i \prod_k a_k}
+ {7 \over 2}  {\Gi \over R_i^2 a_i^2}
+  \sumk \left(
  {6\Gk \over R_k^2 \aik^2}  
+ {2\Gi \over R_i^2 \aik^2}
+ {4 \Gi \over R_i^2 \aik} \sum_{\ell \neq i}  {1 \over \ail}
\right) + \sum_k  \sum_{\ell} {\Gi \over R_i^2 \gik\gil} \non \\
&& 
\!\!\!
+ \sumk \left( 
- {4 \Gk \over R_k^2 \aik\gik}
- { 2 (4 \beta + 1) \Gi \over R_i^2 a_i \aik}
- {4 \Gi \over R_i^2 \aik} \sum_{\ell} {1 \over \gil}
       \right) 
+ {\Gi \over R_i^2} \sum_k \left( -  {1 \over \gik^2} 
                   + { (4 \beta + 1) \over a_i \gik} \right)
\Bigg]  \non \\
&&
\!\!\!\!\!\!\!\!\!\!\!\!\!\!\!\!\!\!\!\!
+\, (1-\de_{ij})
\Bigg[ \frac{\beta-1}{a_i a_j \prod_k a_k}
+ \sum_{k\neq i,j} 
{ \Gk \over R_k^2 }
\left( - {2 \over  \aik} + {1 \over \gik} \right)
\left( - {2 \over  \ajk} + {1 \over \gjk} \right) 
\\
&& 
\!\!\!\!\!\!\!\!\!\!\!\!\!\!\!\!\!\!\!
+ \, \Bigg\{ {\Gi \over R_i^2} 
\left( {2 \over \aij} + {1 \over \gij} \right)
\left(-\sumk {2 \over \aik} + \sum_k {1 \over \gik} 
+ {\left( 2 \beta + \half \right) \over a_i}\right)
+{\Gi \over R_i^2}  \left( - {2 \over \aij^2} - {1\over \gij^2} \right)
 + ( i \leftrightarrow j) \Bigg\} 
\Bigg] \non
\eea
where now 
$\aij= a_i - a_j$, $\gij= a_i + a_j$,  $R_i = \prod_{j \neq i} \aij$,
and $\Gi = a_i^{2\beta} \prod_{j} (a_i + a_j)$.
Equations (\ref{SAtaupert}) and (\ref{SAtauone})
may be written succinctly as the second derivative 
of the prepotential (\ref{SApre}).

\begingroup\raggedright\endgroup

\end{document}